\documentclass[12pt,a4paper,twoside]{article}
\usepackage{latexsym,amsmath,amssymb,amsfonts,amsthm}
\usepackage[mathscr]{eucal}
\usepackage{makeidx}

\def\nabla{\bigtriangledown}
\newcommand{ \R} {\mbox{\rm I$\!$R}}
\newcommand{ \C} {\mbox{\rm I$\!$C}}
\newcommand{ \Z} {\mbox{\rm I$\!$Z}}
\makeindex
\begin{document}

\title{Noncommutative Finsler Geometry, \\
Gauge Fields and Gravity }
\author{Sergiu I. Vacaru \thanks{%
E-mail address:\ vacaru@fisica.ist.utl.pt, ~~
sergiu$_{-}$vacaru@yahoo.com,\
} \\
{\small \textit{Centro Multidisciplinar de Astrofisica - CENTRA,
Departamento de Fisica,}}\\
{\small \textit{Instituto Superior Tecnico, Av. Rovisco Pais 1,
Lisboa,
1049-001, Portugal}}\\
}
\date{\today }
\maketitle

\begin{abstract}
The work extends the A. Connes' noncommutative geometry to spaces
with generic local anisotropy. We apply the E. Cartan's
anholonomic frame approach to geometry models and physical
theories and develop the nonlinear connection formalism for
projective module spaces.  Examples of noncommutative generation
of anholonomic Riemann, Finsler and Lagrange spaces are analyzed.
We also present a research on noncommutative Finsler--gauge
theories, generalized Finsler gravity and anholonomic (pseudo)
Riemann geometry which appear naturally if anholonomic frames
(vierbeins) are defined in the context of string/M--theory and
extra dimension Riemann gravity. \vskip5pt.

Pacs:\ 02.40.Gh, 02.40.-k, 04.50.+h

\end{abstract}

\tableofcontents



\section{Introduction}

In the last twenty years, there has been an increasing interest in
noncommutative and/or quantum geometry with applications both in
mathematical and particle physics. It is now generally considered
that at very high energies, the spacetime can not be described by
a usual manifold structure. Because of quantum fluctuations, it
is difficult to define localized points and the quantum spacetime
structure is supposed to posses generic noncommutative, nonlocal
and locally anisotropic properties. Such ideas originate from the
suggestion that the spacetime coordinates do not commute at a
quantum level \cite{snyder}, they are present in the modern
string theory \cite{deligne,polchinski} and background the
noncommutative physics and geometry \cite{connes} and quantum
geometry \cite{qg}.

Many approaches can be taken to introducing noncommutative
geometry and
developing noncommutative physical theories (Refs. \cite%
{connes,douglas,dubois,gracia,konechny,landi,madore,varilly}
emphasize some basic monographs and reviews). \ This paper has
three aims: \ First of all we would like to give an exposition of
some basic facts on anholonomic frames and associated nonlinear
connection structures both on commutative and noncommutative
spaces (respectively modeled in vector bundles and in projective
modules of finite type). \ Our second goal is to state the
conditions when different variants of Finsler, Lagrange and
generalized Lagrange geometries, in commutative and
noncommutative forms, can be defined by corresponding frame,
metric and connection structures. The third aim is to construct
and analyze properties of gauge and gravitational noncommutative
theories with generic local anisotropy and to prove that such
models can be elaborated in the framework of noncommutative
approaches to Riemannian gravity theories.

This paper does not concern the topic of Finsler like commutative
and noncommutative structres in string/M-theories (see the Ref.
\cite{vfncs}, which can be considered as a string partner of this
work).

We are inspired by the geometrical ideas from a series of
monographs and works by E. Cartan \cite{cartan} where a unified
moving frame approach to the Riemannian and Finsler geometry,
Einstein gravity and Pffaf systems, bundle spaces and spinors, as
well the preliminary ideas on nonlinear connections and various
generalizations of gravity theories were developed. By
considering anholonomic frames on (pseudo) Riemannian manifolds
and in tangent and vector bundles, we can model very sophisticate
geometries with local anisotropy. We shall apply the concepts and
methods developed by the
Romanian school on Finsler geometry and generalizations \cite%
{miron,ma,bejancu,vmon1} from which we leaned that the Finsler
and Cartan like geometries may be modeled on vector (tangent) and
covector (cotangent) bundles if the constructions are adapted to
the corresponding nonlinear connection structure via anholonomic
frames. In this case the geometric ''picture'' and physical
models have a number of common points with those from the usual
Einstein--Cartan theory and/or extra dimension (pseudo)
Riemannian geometry. As general references on Finsler geometry and
applications we cite the monographs \cite%
{finsler,miron,ma,bejancu,vmon1,vmon2}) and point the fact that
the bulk of works on Finsler geometry and generalizations
emphasize differences with the usual Riemannian geometry rather
than try to approach them from a unified viewpoint (as we propose
in this paper).

By applying the formalism of nonlinear connections (in brief,
N--connection) and adapted anholnomic frames in vector bundles
and superbundles we extended the geometry of Clifford structures
and spinors for generalized Finsler
spaces and their higher order extensions in vector--covector bundles \cite%
{vspinors,vmon2}, constructed and analyzed different models of
gauge theories and gauge gravity with generic anisotropy
\cite{vgauge}, defined an anisotropic stochastic calculus in
bundle and superbundle spaces provided with nonlinear connection
structure \cite{vstoch,vmon1}, with a number of applications in
the theory of anisotropic kinetic and thermodynamic processes
\cite{vankin}, developed supersymmetric theories with local
anisotropy \cite{vsuper,vmon1,vstring} and proved that Finsler
like (super)
geometries are contained alternatively in modern string theory \cite%
{vstring,vmon1}. One should be emphasized here that in our
approach we have not proposed any ''exotic'' locally anisotropic
string theories modifications but demonstrated that anisotropic
structures, Finsler like or another ones, may appear
alternatively to the Riemannian geometry, or even can be modeled
in the framework of a such geometry, in the low energy limit of
the string theory, because we are dealing with frame, vierbein,
constructions.

The most surprising fact was that the Finsler like structures
arise in the usual (pseudo) Riemannian geometry of lower and
higher dimensions and even in the Einstein gravity. References
\cite{vexsol} contain investigations of a number of exact
solutions in modern gravity theories (Einstein, Kaluza--Klein and
string/brane gravity) which describe locally anisotropic
wormholes, Taub NUT spaces, black ellipsoid/torus solutions,
solitonic and another type configurations. It was proposed a new
consequent method of constructing exact solutions of the Einstein
equations for off--diagonal metrics, in spaces of dimension
$d>2,$ depending on three and more isotropic and anisotropic
variables which are effectively diagonalized by anholonomic frame
transforms. The vacuum and matter field equations are reduced to
very simplified systems of partial differential equations which
can be integrated in general form \cite{vmethod}.

A subsequent research in Riemann--Finsler and noncommutative
geometry and physics requires the investigation of the fact if
the A. Connes functional analytic approach to noncommutative
geometry and gravity may be such way generalized as to include
the Finsler, and of another type anisotropy, spaces. The first
attempt was made in Refs. \cite{vnonc} where some models of
noncommutative gauge gravity (in the commutative limit being
equivalent to the Einstein gravity, or to different
generalizations to de Sitter, affine, or Poincare gauge gravity
with, or not, nonlinear realization of the gauge groups) were
analyzed. \ Further developments in formulation of noncommutative
geometries with anholonomic and anisotropic structures and their
applications in modern particle physics lead to a rigorous study
of the geometry of anholonomic noncommutative frames with
associated N--connection structure, to which are devoted our
present researches.

The paper has the following structure: in section 2 we present
the necessary definitions and results on the functional approach
to commutative and noncommutative geometry. Section 3 is devoted
to the geometry of vector bundles and theirs noncommutatie
generalizations as finite projective modules. We define the
nonlinear connection in commutative and noncommutative spaces,
introduce locally anisotropic Clifford/spinor structures and
consider the gravity and gauge theories from the viewpoint of
anholonomic frames with associated nonlinear connection
structures. In section 4 we prove that various type of gravity
theories with generic anisotropy, constructed on anholonomic
Riemannian spaces and their Kaluza--Klein and Finsler like
generalizations can be derived from the A. Connes' functional
approach to noncommutative geometry by applying the canonical
triple formalism but extended to vector bundles provided with
nonlinear connection structure. In section 5, we elaborate and
investigate noncommutative gauge like gravity models (which in
different limits contain the standard Einstein's general
relativity and various its anisotropic and gauge
generalizations). The approach holds true also for (pseudo)
Riemannian metrics, but is based on noncommutative extensions of
the frame and connection formalism. This variant is preferred
instead of the usual metric models which seem to be more
difficult to be tackled in the framework of noncommutative
geometry if we are dealing with pseudo--Euclidean signatures and
with complex and/or nonsymmetic metrics. Finally, we present a
discussion and conclusion of the results in section 6.

\section{Commutative and Noncommutative Spaces}

The A. Connes' functional analytic approach \cite{connes} to the
noncommutative topology and geometry is based on the theory of
noncommutative $C^{\ast }$--algebras. Any commutative $C^{\ast
}$--algebra can be realized as the $C^{\ast }$--algebra of
complex valued functions over locally compact Hausdorff space. A
noncommutative $C^{\ast }$--algebra can be thought of as the
algebra of continuous functions on some 'noncommutative
space' (see details in Refs. \ \cite%
{connes,dubois,gracia,landi,madore,varilly}). \

Commutative gauge and gravity theories stem from the notions of
connections (linear and nonlinear), metrics and frames of
references on manifolds and vector bundle spaces. The possibility
of extending such theories to some noncommutative models is based
on the Serre--Swan theorem \cite{swan} stating that there is a
complete equivalence between the category of (smooth) vector
bundles over a smooth compact space (with bundle maps) and the
category of porjective modules of finite type over commutative
algebras and module morphisms. \ Following that theorem, the
space $\Gamma \left( E\right) $ of smooth sections of a vector
bundle $E$ over a compact space is a projective module of finite
type over the algebra $C\left( M\right) $ of
smooth functions over $M$ and any finite projective $C\left( M\right) $%
--module can be realized as the module of sections of some vector
bundle
over $M.$ This construction may be extended if a noncommutative algebra $%
\mathcal{A}$ is taken as the starting ingredient: the
noncommutative
analogue of vector bundles are projective modules of finite type over $%
\mathcal{A}$. This way one developed a theory of linear
connections which culminates in the definition of Yang--Mills
type actions or, by some much more general settings, one
reproduced Lagrangians for the Standard model with its Higgs
sector or different type of gravity and Kaluza--Klein models
(see, for instance, Refs \cite{cl,ch1,landi1,madore}).

This section is devoted to the theory of nonlinear connections in
pojective \ modules of finite type over a noncommutative algebra
$\mathcal{A}$. We shall introduce the basic definitions and
present the main results connected with anhlolonomic frames and
metric structures in such noncommutative spaces.

\subsection{Algebras of functions and (non) commutative spaces}

The general idea of noncommutative geometry is to shift from
spaces to the algebras of functions defined on them. In this
subsection, we give some general facts about algebras of
continuous functions on topological spaces, analyze the concept
of modules as bundles and define the nonlinear connections. We
present mainly the objects we shall need later on while
referring to \cite%
{connes,douglas,dubois,gracia,konechny,landi,madore,varilly} for
details. $\ $

We start with $\ $some necessary definitions on $C^{\ast
}$--algebras and compact operators

In this work any algebra $\mathcal{A}$ is an algebra over the
field of complex numbers $\C$, i. e. $\mathcal{A}$ is a vector
space over $\C$ when
the objects like $\alpha a\pm \beta b,$ with $a,b\in \mathcal{A}$ and $%
\alpha ,\beta \in \C,$ make sense. Also, there is defined (in
general) a noncommutative product $\mathcal{A}$ $\times
\mathcal{A}$ $\rightarrow
\mathcal{A}$ when for every elements $\left( a,b\right) $ and $a,b,$ $%
\mathcal{A}$ $\times \mathcal{A}$ $\ni \left( a,b\right)
\rightarrow ab\in
\mathcal{A}$ the conditions of distributivity,%
\begin{equation*}
a(b+c)=ab+ac,~(a+b)c=ac+bc,
\end{equation*}
for any $a,b,c\in \mathcal{A},$ in general, $ab\neq ba.$ It is
assumed that there is a unity $I\in \mathcal{A}.$

The algebra $\mathcal{A}$ is considered to be a so--called ''$\ast $%
--algebra'', for which an (antilinear) involution $\ast :\mathcal{A}$ $%
\rightarrow \mathcal{A}$ is defined by the properties%
\begin{equation*}
a^{\ast \ast }=a,~\left( ab\right) ^{\ast }=b^{\ast }a^{\ast
},~\left( \alpha a+\beta b\right) ^{\ast }=\overline{\alpha
}a^{\ast }+\overline{\beta }b^{\ast },
\end{equation*}%
where the bar operation denotes the usual complex conjugation.

One also considers $\mathcal{A}$ to be a normed algebra with a norm \ $%
\left| \left| \cdot \right| \right| $ $:\mathcal{A}$ $\rightarrow
\R,$ where
$\R$ the real number field, satisfying the properties%
\begin{eqnarray*}
\left| \left| \alpha a\right| \right| &=&|\alpha |\left| \left|
a\right| \right| ;~\left| \left| a\right| \right| \geq 0,~\left|
\left| a\right|
\right| =0\Leftrightarrow a=0; \\
\left| \left| a+b\right| \right| &\leq &\left| \left| a\right|
\right| +\left| \left| b\right| \right| ;~\left| \left| ab\right|
\right| \leq \left| \left| a\right| \right| \left| \left|
b\right| \right| .
\end{eqnarray*}%
This allows to define the 'norm' or 'uniform' topology when an $\varepsilon $%
--neighborhood of any $a\in \mathcal{A}$ is given by%
\begin{equation*}
U\left( a,\varepsilon \right) =\left\{ b\in \mathcal{A},\left|
\left| a-b\right| \right| <\varepsilon \right\} ,\varepsilon >0.
\end{equation*}

A Banach algebra is a normed algebra which is complete in the
uniform topolgy and a Banach $\ast $--algebra is a normed $\ast
$--algebra which is complete and such that $\left| \left| a^{\ast
}\right| \right| =\left|
\left| a\right| \right| $ for every $a\in \mathcal{A}.$ We can define now a $%
C^{\ast }$--algebra $\mathcal{A}$ as a Banach $\ast $--algebra
with the norm satisfying the additional identity $\left| \left|
a^{\ast }a\right| \right| =\left| \left| a\right| \right| ^{2}$
for every $a\in \mathcal{A}.$

We shall use different commutative and noncommutative algebras:

By $\mathcal{C}(M)$ one denotes the algebra of continuous
functions on a compact Hausdorf topological space $M,$ with $\ast
$ treated as the complex conjugation and the norm given by the
supremum norm, $||f||_{\infty }=\sup_{x\in M}|f(x)|.$ If the
space $M$ is only locally compact, one writes
$\mathcal{C}_{0}(M)$ for the algebra of continuous functions
vanishing at infinity (this algebra has no unit).

The $\mathcal{B(H)}$ is used for the noncommutative algebra of
bounded operators on an infinite dimensional Hilbert space
$\mathcal{H}$ with the involution $\ast $ given by the adjoint
and the norm defined as the operator
norm%
\begin{equation*}
||A||=\sup \left\{ ||A\zeta ||;\zeta \in \mathcal{H},~A\in \mathcal{B(H)}%
,~||\zeta ||\leq 1\right\} .
\end{equation*}

One considers the noncommutative algebra $M_{n}\left( \C\right) $ of $%
n\times n$ matrices $T$ with complex entries, when $T^{\ast }$ is
considered
as the Hermitian conjugate of $T.$ We may define a norm as%
\begin{equation*}
||T||=\{\mbox{the positive square root of the largest eigenvalue
of }\ T^{\ast }T\}
\end{equation*}%
or as
\begin{equation*}
||T||^{\prime }=\sup [T_{ij}],~T=\{T_{ij}\}.
\end{equation*}%
The last definition does not define a $C^{\ast }$--norm, but both
norms are
equivalent as Banach norm because they define the same topology on $%
M_{n}\left( \C\right) .$

A left (right) ideal $\mathcal{T}$ is a subalgebra $\mathcal{A}\in \mathcal{T%
}$ $\ $\ if $a\in \mathcal{A}$ and $b\in \mathcal{T}$ $\ $imply
that $ab\in \mathcal{T}$ $\ (ba\in \mathcal{T}).$ A two sided
ideal is a subalgebra
(subspace) which is both a left and right ideal. An ideal $\mathcal{T}$ $\ $%
\ is called maximal if there is not other ideal of the same type
which contain it. For a Banach $\ast $--algebra $\mathcal{A}$ and
two--sided $\ast
$--ideal $\mathcal{T}$ (which is closed in the norm topology) we can make $%
\mathcal{A}/\mathcal{T}$ $\ $\ a Banach $\ast $--algebra. This
allows to
define the quotient $\mathcal{A}/\mathcal{T}$ $\ $to be a $C^{\ast }$%
--algebra if $\mathcal{A}$ \ is a $C^{\ast }$--algebra. A $C^{\ast }$%
--algebra is called simple if it has no nontrivial two--sided
ideals. A two--sided ideal is called essential in a $C^{\ast
}$--algebra if any other non--zero ideal in this algebra has a
non--zero intersection with it.

One \ defines the resolvent set $r(a)$ of an element $a\in
\mathcal{A}$ as a the subset of complex numbers given by
$r(a)=\{\lambda \in \C|a-\lambda I$ is invertible\}. The
resolvent of $a$ at any $\lambda \in $ $r(a)$ is given by the
inverse $\left( a-\lambda I\right) ^{-1}.$ The spectrum $\sigma
\left( a\right) $ of an element $a$ is introduced as the
complement of $r(a)$ $\ $in $\C.$ For $C^{\ast }$--algebras the
spectrum of any element \ is a nonempty compact subset of $\C.$
The spectral radius $\rho \left( a\right) $ of $a\in \mathcal{A}$
$\ $\ is defined $\rho \left( a\right) =\sup
\{|\lambda |,\lambda \in r(a)\};$ for $\mathcal{A}$ being a $C^{\ast }$%
--algebra, one obtains $\rho \left( a\right) =||a||$ for every
$a\in \mathcal{A}.$ This distinguishes the $C^{\ast }$--algebras
as those for which the norm may be uniquely determined by the
algebraic structure. One considers self--adjoint elements for
which $a=a^{\ast },$ such elements have real spectra and satisfy
the conditions $\sigma (a)\subseteq \left[ -||a||,||a||\right] $
and $\sigma (a^{2})\subseteq \left[ 0|,||a||\right] .$ An element
$a$ is positive, $\ $i. e. $a>0,$ if its spectrum belongs to the
positive half--line. This is possible if and only if $a=bb^{\ast
}$ for some $b\in \mathcal{A}.$

One may consider $\ast $--morphisms between two $C^{\ast }$--algebras $%
\mathcal{A}$ and $\mathcal{B}$ as some $\C$--linear maps $\pi :\mathcal{A}$ $%
\rightarrow $ $\mathcal{B}$ which are subjected to the additional conditions%
\begin{equation*}
\pi (ab)=\pi (a)\pi (b),~\pi (a^{\ast })=\pi (a)^{\ast }
\end{equation*}%
which imply that $\pi $ are positive and continuous and that $\pi (\mathcal{A%
})$ is a $C^{\ast }$--subalgebra of $\mathcal{B}$ (see, for instance, \cite%
{landi}). We note that a $\ast $--morphism which is bijective as
a map defines a $\ast $--isomorphism for which the inverse map
$\pi ^{-1}$ is automatically a $\ast $--morphism.

In order to construct models of noncommutative geometry one uses
representations of a $C^{\ast }$--algebra $\mathcal{A}$ as pairs
$\left( \mathcal{H},\pi \right) $ where $\mathcal{H}$ is a
Hilbert space and $\pi $
is a $\ast $--morphism $\pi :$ $\mathcal{A\rightarrow B}\left( \mathcal{H}%
\right) $ with $\mathcal{B}\left( \mathcal{H}\right) $ being the $C^{\ast }$%
--algebra of bounded operators on $\mathcal{H}.$ There are
different
particular cases of representations: A representation $\left( \mathcal{H}%
,\pi \right) $ is faithful if $\ker \pi =\{0\},$ i. e. $\pi $ is a $\ast $%
--isomorphism between $\mathcal{A}$ and $\pi (\mathcal{A})$ which
holds if and only if $||\pi (a)||=||a||$ for any $a\in
\mathcal{A}$ or $\pi (a)>0$ for all $a>0.$ A representation is
irreducible if the only closed subspaces of $\mathcal{H}$ which
are invariant under the action of $\pi (\mathcal{A})$ are the
trivial subspaces $\{0\}$ and $\mathcal{H}.$ It can be proven that
if the set of the elements in $\mathcal{B}\left(
\mathcal{H}\right) $ commute with each element in $\pi
(\mathcal{A}),$ i. e. the set consists of multiples of the
identity operator, the representation is irreducible. Here we
note that two representations $\left( \mathcal{H}_{1},\pi
_{1}\right) $ and $\left( \mathcal{H}_{2},\pi _{2}\right) $ are
said to be (unitary) equivalent if there exists a unitary
operator $U:\mathcal{H}_{1}\rightarrow
\mathcal{H}_{2}$ such that $\pi _{1}(a)=U^{\ast }\pi _{2}(a)U$ for every $%
a\in \mathcal{A}.$

A subspace (subalgebra) $\mathcal{T}$ of the $C^{\ast }$--algebra $\mathcal{A%
}$ is a primitive ideal if $\mathcal{T}=\ker \pi $ for some
irreducible representation $\left( \mathcal{H},\pi \right) $ of
$\mathcal{A}.$ In this case $\mathcal{T}$ $\ $is automatically a
closed two--sided ideal. One say that $\mathcal{A}$ is a
primitive $C^{\ast }$--algebra if $\mathcal{A}$ has a faithful
irreducible representation on some Hilbert space for which the
set $\{0\}$ is a primitive ideal. One denotes by $\Pr
im\mathcal{A}$ the set of all primitive ideals of a $C^{\ast
}$--algebra $\mathcal{A}.$

Now we recall some basic definitions and properties of compact
operators on Hilbert spaces \cite{reed}:

Let us first consider the class of operators which may be thought
as some infinite dimensional matrices acting on an infinite
dimensional Hilbert
space $\mathcal{H}.$ More exactly, an operator on the Hilbert space $%
\mathcal{H}$ is said to be of finite rank if the orthogonal
component of its null space is finite dimensional. An operator
$T$ on $\mathcal{H}$ which can be approximated in norm by finite
rank operators is called compact. It can be characterized by the
property that for every $\varepsilon >0$ there is a finite
dimensional subspace $E\subset \mathcal{H}:||T_{|E^{\perp
}}||<\varepsilon ,$ where the orthogonal subspace $E^{\perp }$ is
of finite
codimension in $\mathcal{H}.$ $\ $This way we may define the set $\mathcal{%
K(H)}$ of \ all compact operators on the Hilbert spaces which is
the largest
two--sided ideal in the $C^{\ast }$--algebra $\mathcal{B}\left( \mathcal{H}%
\right) $ of all bounded operators. This set is also a $C^{\ast
}$--algebra with no unit, since the operator $I$ on an infinite
dimensional Hilbert
space is not compact, it is the only norm closed and two--sided when $%
\mathcal{H}$ is separable. We note that the defining representation of $%
\mathcal{K(H)}$ by itself is irreducible and it is the only
irreducible representation up to equivalence.

For an arbitrary $C^{\ast }$--algebra $\mathcal{A}$ acting
irreducibly on a
Hilbert space $\mathcal{H}$ and having non--zero intersection with $\mathcal{%
K(H)}$ one holds $\mathcal{K(H)}$ $\subseteq \mathcal{A}.$ In the
particular
case of finite dimensional Hilbert spaces, for instance, for $\mathcal{H}=\C%
^{n},$ we may write $\mathcal{B}\left( \C^{n}\right) =\mathcal{K}(\C%
^{n})=M_{n}\left( \C\right) ,$ which is the algebra of $n\times
n$ matrices with complex entries. Such algebra has only one
irreducible representation (the defining one).

\subsection{Commutative spaces}

Let us denote by $\mathcal{C}$ a fixed commutative $C^{\ast
}$--algebra with unit and by $\widehat{\mathcal{C}}$ the
corresponding structure space defined \ as the space of
equivalence classes of irreducible representations of
$\mathcal{C}$ ( $\widehat{\mathcal{C}}$ does not contains the
trivial representation $\mathcal{C}$ $\rightarrow \{0\}).$ One
can define a
non--trivial $\ast $--linear multiplicative functional $\phi :\mathcal{C}%
\rightarrow \C$ with the property that $\phi \left( ab\right)
=\phi \left(
a\right) \phi \left( b\right) $ for any $a$ and $b$ from $\mathcal{C}$ and $%
\phi (I)=1$ for every $\phi \in $ $\widehat{\mathcal{C}}.$ Every
such
multiplicative functional defines a character of $\mathcal{C},$ i. e. $%
\widehat{\mathcal{C}}$ is also the space of all characters of
$\mathcal{C}.$

The Gel'fand topology \ is the one with point wise convergence on $\mathcal{C%
}.$ A sequence $\left\{ \phi _{\varpi }\right\} _{\varpi \in \Xi
}$ of elements of $\widehat{\mathcal{C}}$, where \ $\Xi $ is any
directed set, \
converges to $\phi (c)\in \widehat{\mathcal{C}}$ if and only if for any $%
c\in \mathcal{C},$ the sequence $\left\{ \phi _{\varpi
}(c)\right\} _{\varpi
\in \Xi }$ converges to $\phi (c)$ in the topology of $\C.$ If the algebra $%
\mathcal{C}$ has a unite, $\widehat{\mathcal{C}}$ is a compact
Hausdorff space (a topoligical space is Hausdorff if for any two
points of the space there are two open disjoint neighborhoods
each containing one of the point,
see Ref. \cite{kel}). The space $\widehat{\mathcal{C}}$ is only compact if $%
\mathcal{C}$ is without unit. This way the space
$\widehat{\mathcal{C}}$ (called the Gel'fand space) is made a
topological space. We may also consider $\widehat{\mathcal{C}}$
$\ $as a space of maximal ideals, two sided, of $\mathcal{C}$ $\
$instead of the space of irreducible representations. If there is
no unit, the ideals to be considered should be regular (modular),
see details in Ref. \cite{dix}. Considering $\phi \in \C,$
we can decompose $\mathcal{C}=Ker\left( \phi \right) \oplus \C,$ where $%
Ker\left( \phi \right) $ is an ideal of codimension one and so is
a maximal
ideal of $\mathcal{C}.$ Considered in terms of maximal ideals, the space $%
\widehat{\mathcal{C}}$ is given the Jacobson topology,
equivalently, hull kernel topology (see next subsection for
general definitions for both commutative and noncommutative
spaces), producing a space which is homeomorphic to the one
constructed by means of the Gel'fand topology.

Let us consider an example when the algebra $\mathcal{C}$
generated by $s$
commuting self--adjoint elements $x_{1},...x_{s}.$ The structure space $%
\widehat{\mathcal{C}}$ can be identified with a compact subset of
$\R^{s}$ by the map $\phi (c)\in \widehat{\mathcal{C}}\rightarrow
\left[ \phi (x_{1}),...,\phi (x_{s})\right] \in \R^{s}.$ This map
has a joint spectrum of $x_{1},...x_{s}$ as the set of all
$s$--tuples of eigenvalues corresponding to common eigenvectors.

In general, we get an interpretation of elements $\mathcal{C}$ as $\C$%
--valued continuous functions on $\widehat{\mathcal{C}}.$ The
Gel'fand--Naimark theorem (see, for instance, \cite{dix}) states
that all
continuous functions on $\widehat{\mathcal{C}}$ are of the form $\widehat{c}%
(\phi )=\phi \left( c\right) ,$ which defines the so--called
Gel'fand
transform for every $\phi (c)\in \widehat{\mathcal{C}}$ and the map $%
\widehat{c}:$ $\widehat{\mathcal{C}}\rightarrow \C$ being
continuous for
each $c.$ A transform $c\rightarrow \widehat{c}$ is isometric for every $%
c\in \mathcal{C}$ if $||\widehat{c}||_{\infty }=$ $||c||,$ with $%
||...||_{\infty }$ defined at the supremum norm on
$\mathcal{C}\left( \widehat{\mathcal{C}}\right) .$

The Gel'fand transform can be extended for an arbitrary locally
compact
topological space $M$ for which there exists a natural $C^{\ast }$--algebra $%
\mathcal{C}$ $(M).$ On can be identified both set wise and
topologically the Gel'fand space $\widehat{\mathcal{C}}(M)$ \ and
the space $M$ itself through
the evaluation map%
\begin{equation*}
\phi _{x}:\mathcal{C}(M)\rightarrow \C,~\phi _{x}(f)=f(x)
\end{equation*}%
for each $x\in M,$ where $\phi _{x}\in \widehat{\mathcal{C}}(M)$
gives a complex homomorphism. Denoting by $\mathcal{I}_{x}=\ker
\phi _{x},$ which is the maximal ideal of $\mathcal{C}$ $(M)$
consisting of all functions vanishing at $x,$ one proves
\cite{dix} that the map $\phi _{x}$ is a homomorphism of $M$ onto
$\widehat{\mathcal{C}}(M),$ and, equivalently, every maximal
ideal of $\mathcal{C}$ $(M)$ is of the form $\mathcal{I}_{x}$ for
some $x\in M.$

We conclude this subsection: There is a one--to--one
correspondence between the $\ast $--isomorphism classes of
commutative $C^{\ast }$--algebras and the homomorphism classes of
locally compact Hausdorff spaces (such commutative $C^{\ast
}$--algebras with unit correspond to compact Hausdorff spaces).
This correspondence defines a complete duality between the
category of (locally) compact Hausdorff spaces and (proper, when
a map $\ f$ relating two locally compact Hausdorff spaces
$f:X\rightarrow Y$ has the property that $f^{-1}\left( K\right) $
is a compact subset of $X$ when $K$ is a compact subset of $Y,$
and ) continuous maps and the category of commutative (non
necessarily) unital $C^{\ast }$--algebras and $\ast
$--homomorphisms.
In result, any commutative $C^{\ast }$--algebra can be realized as the $%
C^{\ast }$--algebra of complex valued functions over a (locally)
compact Hausdorff space. It should be mentioned that the space
$M$ is a metrizable topological space, i. e. its topology comes
from a metric, if and only if the $C^{\ast }$--algebra is norm
separable (it admits a dense in norm countable subset). This
space is connected topologically if the corresponding algebra has
no projectors which are self--adjoint, $p^{\ast }=p $ and satisfy
the idempotentity condition $p^{2}=p.$

We emphasize that the constructions considered for commutative
algebras cannot be directly generalized for noncommutative
$C^{\ast }$--algebras.

\subsection{Noncommutative spaces}

For a given noncommutative $C^{\ast }$--algebra, there is more
than one
candidate for the analogue of the topological space $M.$ Following Ref. \cite%
{landi} (see there the proofs of results and Appendices), we
consider two possibilities:

\begin{itemize}
\item To use the space $\widehat{\mathcal{A}}$ , \ called the structure
space of $\ $the noncommutative $C^{\ast }$--algebra
$\mathcal{A},$ which is
the space of all unitary equivalence classes of irreducible $\ast $%
--representations.

\item To use the space $\Pr im\mathcal{A},$ called the primitive spectrum of
$\mathcal{A},$ which is the space of kernels of irreducible $\ast $%
--representations (any element of $\Pr im\mathcal{A}$ is
automatically a two--sided $\ast $--ideal of $\mathcal{A)}$.
\end{itemize}

The spaces $\widehat{\mathcal{A}}$ and $\Pr im\mathcal{A}$ agree
for a commutative $C^{\ast }$--algebra, for instance,
$\widehat{\mathcal{A}}$ may be very complicate while $\Pr
im\mathcal{A}$ consisting of a single point.

Let us examine a simple example of generalization to
noncommutative $C^{\ast }$--algebra given by the $2\times 2$
complex matrix algebra
\begin{equation*}
M_{2}(\C)=\{\left[
\begin{array}{cc}
a_{11} & a_{12} \\
a_{21} & a_{22}%
\end{array}%
\right] ,~a_{ij}\in \C\}.
\end{equation*}%
The commutative subalgebra of diagonal matrices
$\mathcal{C}=\{diag[\lambda _{1},\lambda _{2}],~\lambda _{1,2}\in
\C\}$ has a structure space consisting of two points given by the
characters $\ \phi _{1,2}(\left[
\begin{array}{cc}
\lambda _{1} & 0 \\
0 & \lambda _{2}%
\end{array}%
\right] )=\lambda _{1,2}.$ These two characters extend as pure
states to the
full algebra $M_{2}(\C)$ by the maps $\widetilde{\phi }_{1,2}:M_{2}(\C%
)\rightarrow \C,$%
\begin{equation*}
\widetilde{\phi }_{1}\left( \left[
\begin{array}{cc}
a_{11} & a_{12} \\
a_{21} & a_{22}%
\end{array}%
\right] \right) =a_{11},~\widetilde{\phi }_{2}\left( \left[
\begin{array}{cc}
a_{11} & a_{12} \\
a_{21} & a_{22}%
\end{array}%
\right] \right) =a_{22}.
\end{equation*}%
Further details are given in Appendix B to Ref. \cite{landi}.

It is possible to define natural topologies on $\widehat{\mathcal{A}}$ and $%
\Pr im\mathcal{A},$ for instance, by means of a closure
operation. \ For a subset $Q\subset \Pr im\mathcal{A},$ the
closure $\overline{Q}$ is by definition the subset of all
elements in $\Pr im\mathcal{A}$ containing the intersection $\cap
Q$ of the elements of $Q,~\overline{Q}\doteqdot \left\{
\mathcal{I}\in \Pr im\mathcal{A}:\cap Q\subseteq
\mathcal{I}\right\} .$ It is possible to check that such subsets
satisfy the Kuratowski topology axioms and this way defined
topology on $\Pr im\mathcal{A}$ is called the Jacobson topology
or hull--kernel topology, for which $\cap Q$ is the kernel of $Q$
and $~\overline{Q}$ is the hull of $\cap Q$ (see \cite{landi,dix}
on the properties of this type topological spaces).

\section{Nonlinear Connections in Noncommutative Spa\-ces}

In this subsection we define the nonlinear connections in module
spaces, i. e. in noncommutative spaces. The concept on nonlinear
connection came from Finsler geometry (as a set of coefficients
it is present in the works of E. Cartan \cite{cartan}, then the
concept was elaborated in a more explicit fashion by A. Kawaguchi
\cite{kaw}). The global formulation in commutative spaces is due
to W. Barthel \ \cite{barthel} and it was developed in details
for vector, covector and higher order bundles
\cite{ma,miron,bejancu},
spinor bundles \cite{vspinors,vmon2}, superspaces and superstrings \cite%
{vsuper,vmon1,vstring} and in the theory of exact off--diagonal
solutions of the Einstein equations \cite{vexsol,vmethod}. The
concept of nonlinear connection can be extended in a similar
manner from commutative to noncommutative spaces if a
differential calculus is fixed on a noncommutative vector (or
covector) bundle.

\subsection{Modules as bundles}

A vector bundle $E\rightarrow M$ over a manifold $M$ is completely
characterized by the space $\mathcal{E}=\Gamma \left( E,M\right)
$ over its smooth sections defined as a (right) module over the
algebra of $C^{\infty }\left( M\right) $ of smooth functions over
$M.$ It is known the Serre--Swan theorem \cite{swan} which states
that locally trivial, finite--dimensional complex vector bundles
over a compact Hausdorff space $M$ correspond
canonically to finite projective modules over the algebra \ $\mathcal{A}%
=C^{\infty }\left( M\right).$ Inversely, for $\mathcal{E}$ being
a finite projective modules over $C^{\infty }\left( M\right) ,$
the fiber $E_{m}$ of
the associated bundle $E$ over the point $x\in M$ is the space $E_{x}=%
\mathcal{E}/\mathcal{EI}_{x}$ where the ideal is given by%
\begin{equation*}
\mathcal{I}_{x}=\ker \{\xi _{x}:C^{\infty }\left( M\right) \rightarrow \C%
;\xi _{x}(x)=f(x)\}=\{f\in C^{\infty }\left( M\right) |f(x)=0\}\in \mathcal{C%
}\left( M\right) .
\end{equation*}%
If the algebra $\mathcal{A}$ is taken to play the role of smooth
functions
on a noncomutative, instead of the commutative algebra smooth functions $%
C^{\infty }\left( M\right) $, the analogue of a vector bundle is
provided by a projective module of finite type (equivalently,
finite projective module) over $\mathcal{A}.$ On considers the
proper construction of projective modules of finite type
generalizing the Hermitian bundles as well the notion of Hilbert
module when $\mathcal{A}$ is a $C^{\ast }$--algebra in the
Appendix C of Ref. \cite{landi}.

A vector space $\mathcal{E}$ over the complex number field $\C$
can be defined also as a right module of an algebra $\mathcal{A}$
over $\C$ \ which carries a right representation of
$\mathcal{A},$ when for every map of elements $\mathcal{E}$
$\times \mathcal{A}\ni \left( \eta ,a\right) \rightarrow \eta
a\in \mathcal{E}$ one hold the properties
\begin{equation*}
\lambda (ab)=(\lambda a)b,~\lambda (a+b)=\lambda a+\lambda
b,~(\lambda +\mu )a=\lambda a+\mu a
\end{equation*}%
fro every $\lambda ,\mu \in \mathcal{E}$ and $a,b\in \mathcal{A}.$

Having two $\mathcal{A}$--modules $\mathcal{E}$ and
$\mathcal{F},$ a
morphism of $\mathcal{E}$ into $\mathcal{F}$ is \ any linear map $\rho :%
\mathcal{E}$ $\rightarrow $ $\mathcal{F}$ $\ $which is also $\mathcal{A}$%
--linear, i. e. $\rho (\eta a)=\rho (\eta )a$ for every $\eta \in
\mathcal{E} $ and $a\in \mathcal{A}.$

We can define in a similar (dual) manner the left modules and
theirs morphisms which are distinct from the right ones for
noncommutative algebras
$\mathcal{A}.$ A bimodule over an algebra $\mathcal{A}$ is a vector space $%
\mathcal{E}$ which carries both a left and right module
structures. We may define the opposite algebra $\mathcal{A}^{o}$
with elements $a^{o}$ being in bijective correspondence with the
elements \ $a\in \mathcal{A}$ while the multiplication is given
by $\mathcal{\,}a^{o}b^{o}=\left( ba\right) ^{o}.$A right
(respectively, left) $\mathcal{A}$--module $\mathcal{E}$ is
connected
to a left (respectively right) $\mathcal{A}^{o}$--module via relations $%
a^{o}\eta =\eta a^{o}$ (respectively, $a\eta =\eta a).$

One introduces the enveloping algebra $\mathcal{A}^{\varepsilon }=\mathcal{A}%
\otimes _{\C}\mathcal{A}^{o};$ any $\mathcal{A}$--bimodule
$\mathcal{E}$ can be regarded as a right [left]
$\mathcal{A}^{\varepsilon }$--module by setting $\eta \left(
a\otimes b^{o}\right) =b\eta a$ $\quad \left[ \left( a\otimes
b^{o}\right) \eta =a\eta b\right] .$

For a (for instance, right) module $\mathcal{E}$ , we may
introduce a family of elements $\left( e_{t}\right) _{t\in T}$
parametrized by any (finite or infinite) directed set $T$ for
which any element $\eta \in \mathcal{E}$ is expressed as a
combination (in general, in \ more than one manner) $\eta
=\sum\nolimits_{t\in T}e_{t}a_{t}$ with $a_{t}\in \mathcal{A}$
and only a finite number of non vanishing terms in the sum. A
family $\left( e_{t}\right) _{t\in T}$ is free if it consists
from linearly independent elements and defines a basis if any
element $\eta \in \mathcal{E}$ can be written as a unique
combination (sum). One says a module to be free if it admits a
basis. The module $\mathcal{E}$ is said to be of finite type if \
it is finitely generated, i. e. it admits a generating family of
finite cardinality.

Let us consider the module $\mathcal{A}^{K}\doteqdot \C^{K}\otimes _{\C}%
\mathcal{A}.$ The elements of this module can be thought as
$K$--dimensional vectors with entries in $\mathcal{A}$ and
written uniquely as a linear combination $\eta
=\sum\nolimits_{t=1}^{K}e_{t}a_{t}$ were the basis $e_{t}$
identified with the canonical basis of $\C^{K}.$ This is a free
and finite type module. In general, we can have bases of different
cardinality. However, if a module $\mathcal{E}$ \ is of finite
type there is always an integer $K$ and a module surjection $\rho
:\mathcal{A}^{K}\rightarrow \mathcal{E}$ with a base being a
image of a free basis, $\epsilon _{j}=\rho (e_{j});j=1,2,...,K.$

In general, it is not possible to solve the constraints among the
basis
elements as to get a free basis. \ The simplest example is to take a sphere $%
S^{2}$ and the Lie algebra of smooth vector fields on it, $\mathcal{G}=%
\mathcal{G}(S^{2})$ which is a module of finite type over
$C^{\infty }\left( S^{2}\right) ,$ with the basis defined by
$X_{i}=\sum_{j,k=1}^{3}\varepsilon _{ijk}x_{k}\partial /\partial
x^{k};i,j,k=1,2,3,$ and coordinates $x_{i}$ such that
$\sum_{j=1}^{3}x_{j}^{2}=1.$ The introduced basis is not free
because $\sum_{j=1}^{3}x_{j}X_{j}=0;$ there are not global vector field on $%
S^{2}$ which could form a basis of $\mathcal{G}(S^{2}).$ This
means that the tangent bundle $TS^{2}$ is not trivial.

We say that a right $\mathcal{A}$--module $\mathcal{E}$ is
projective if for
every surjective module morphism $\rho :\mathcal{M}$ $\rightarrow $ $%
\mathcal{N}$ splits, i. e. there exists a module morphism \ $s:\mathcal{E}$ $%
\rightarrow $ $\mathcal{M}$ such that $\rho \circ
s=id_{\mathcal{E}}.$ There are different definitions of
porjective modules (see Ref. \cite{landi} on properties of such
modules). Here we note the property that if a $\mathcal{A}
$--module $\mathcal{E}$ is projective, there exists a free module $\mathcal{F%
}$ and a module $\mathcal{E}^{\prime }$ (being a priory
projective) such that $\mathcal{F}=\mathcal{E}\oplus
\mathcal{E}^{\prime }.$

For the right $\mathcal{A}$--module $\mathcal{E}$ being
projective and of finite type with surjection $\rho
:\mathcal{A}^{K}\rightarrow \mathcal{E}$ and following the
projective property we can find a lift $\widetilde{\lambda
}:\mathcal{E}$ $\rightarrow $ $\mathcal{A}^{K}$ such that $\rho
\circ \widetilde{\lambda }=id_{\mathcal{E}}.$ There is a proof of
the property that the module $\mathcal{E}$ is projective of
finite type over $\mathcal{A}$
if and only if there exists an idempotent $p\in End_{\mathcal{A}}\mathcal{A}%
^{K}=M_{K}(\mathcal{A}),$ $p^{2}=p,$ the $M_{K}(\mathcal{A})$
denoting the
algebra of $K\times K$ matrices with entry in $\mathcal{A},$ such that $%
\mathcal{E}=p\mathcal{A}^{K}.$ We may associate the elements of
$\mathcal{E}$ to $K$--dimensional column vectors whose elements
are in $\mathcal{A},$ the collection \ of which are invariant
under the map $p,$ $\mathcal{E}$ $=\{\xi =(\xi _{1},...,\xi
_{K});\xi _{j}\in \mathcal{A},~p\xi =\xi \}.$ For simplicity, we
shall use the term finite projective to mean projective of finite
type.

The noncommutative variant of the theory of vector bundles may be
constructed by using the Serre and Swan theorem \cite{swan,landi}
which states that for a compact finite dimensional manifold $M,$
a $C^{\infty }\left( M\right) $--module $\mathcal{E}$ is
isomorphic to a module $\Gamma \left( E,M\right) $ of smooth
sections of a bundle $E\rightarrow M,$ if and only if it is
finite projective. If $E$ is a complex vector bundle over a
compact manifold $M$ of dimension $n,$ there exists a finite cover $%
\{U_{i},i=1,...,n\}$ of $M$ such that $E_{|U_{i}}$ is trivial.
Thus, the integer $K$ which determines the rank of the free
bundle from which to project onto sections of the bundle is
determined by the equality $N=mn$ where $m$ is the rank of the
bundle (i. e. of the fiber) and $n$ is the dimension of $M.$

\subsection{The commutative nonlinear connection geometry}

Let us remember the definition and the main results on nonlinear
connections in commutative vector bundles as in Ref. \cite{ma}.

\subsubsection{Vector bundles, Riemannian spaces and nonlinear connections}

We consider a vector bundle $\xi =\left( E,\mu ,M\right) $ whose
fibre is $\R
$$^{m}$ and $\mu ^{T}:TE\rightarrow TM$ denotes the differential of the map $%
\mu :E\rightarrow M.$ The map $\mu ^{T}$ is a fibre--preserving
morphism of the tangent bundle $\left( TE,\tau _{E},E\right) $ to
$E$ and of tangent bundle $\left( TM,\tau ,M\right) $ to $M.$ The
kernel of the morphism $\mu ^{T}$ is a vector subbundle of the
vector bundle $\left( TE,\tau _{E},E\right).$ This kernel is
denoted $\left( VE,\tau _{V},E\right) $ and called the vertical
subbundle over $E.$ We denote by $i:VE\rightarrow TE$
the inclusion mapping and the local coordinates of a point $u\in E$ by $%
u^{\alpha }=\left( x^{i},y^{a}\right) ,$ where indices
$i,j,k,...=1,2,...,n$ and $a,b,c,...=1,2,...,m.$

A vector $X_{u}\in TE,$ tangent in the point $u\in E,$ is locally
represented $\left( x,y,X,\widetilde{X}\right) =\left(
x^{i},y^{a},X^{i},X^{a}\right) ,$ where $\left( X^{i}\right) \in
$$\R$$^{n}$
and $\left( X^{a}\right) \in $$\R$$^{m}$ are defined by the equality $%
X_{u}=X^{i}\partial _{i}+X^{a}\partial _{a}$ [$\partial _{\alpha
}=\left(
\partial _{i},\partial _{a}\right) $ are usual partial derivatives on
respective coordinates $x^{i}$ and $y^{a}$]. For instance, $\mu
^{T}\left( x,y,X,\widetilde{X}\right) =\left( x,X\right) $ and
the submanifold $VE$ contains elements of type $\left(
x,y,0,\widetilde{X}\right) $ and the local fibers of the vertical
subbundle are isomorphic to $\R$$^{m}.$ Having $\mu ^{T}\left(
\partial _{a}\right) =0,$ one comes out that $\partial _{a}$ is a
local basis of the vertical distribution $u\rightarrow V_{u}E$ on
$E,$ which is an integrable distribution.

A nonlinear connection (in brief, N--connection) in the vector
bundle $\xi =\left( E,\mu ,M\right) $ is the splitting on the
left of the exact sequence
\begin{equation*}
0\rightarrow VE\rightarrow TE/VE\rightarrow 0,
\end{equation*}%
i. e. a morphism of vector bundles $N:TE\rightarrow VE$ such that
$C\circ i$ is the identity on $VE.$

The kernel of the morphism $N$ is a vector subbundle of $\left(
TE,\tau
_{E},E\right) ,$ it is called the horizontal subbundle and denoted by $%
\left( HE,\tau _{H},E\right) .$ Every vector bundle $\left(
TE,\tau _{E},E\right) $ provided with a N--connection structure
is Whitney sum of the vertical and horizontal subbundles, i. e.
\begin{equation}
TE=HE\oplus VE.  \label{wihit}
\end{equation}
It is proven that for every vector bundle $\xi =\left( E,\mu
,M\right) $ over a compact manifold $M$ there exists a nonlinear
connection \cite{ma}.

Locally a N--connection $N$ is parametrized by a set of
coefficients\newline $N_{i}^{a}(u^{\alpha
})=N_{i}^{a}(x^{j},y^{b})$ which transforms as
\begin{equation*}
N_{i^{\prime }}^{a^{\prime }}\frac{\partial x^{i^{\prime }}}{\partial x^{i}}%
=M_{a}^{a^{\prime }}N_{i}^{a}-\frac{\partial M_{a}^{a^{\prime
}}}{\partial x^{i}}y^{a}
\end{equation*}%
under coordinate transforms on the vector bundle $\xi =\left(
E,\mu
,M\right) ,$%
\begin{equation*}
x^{i^{\prime }}=x^{i^{\prime }}\left( x^{i}\right) \mbox{ and
}y^{a^{\prime }}=M_{a}^{a^{\prime }}(x)y^{a}.
\end{equation*}

If a N--connection structure is defined on $\xi ,$ the operators
of local partial derivatives $\partial _{\alpha }=\left( \partial
_{i},\partial _{a}\right) $ and differentials $d^{\alpha
}=du^{\alpha }=\left( d^{i}=dx^{i},d^{a}=dy^{a}\right) $ should
be elongated as to adapt the local basis (and dual basis)
structure to the Whitney decomposition of the vector
bundle into vertical and horizontal subbundles, (\ref{wihit}):%
\begin{eqnarray}
\partial _{\alpha } &=&\left( \partial _{i},\partial _{a}\right) \rightarrow
\delta _{\alpha }=\left( \delta _{i}=\partial
_{i}-N_{i}^{b}\partial
_{b},\partial _{a}\right) ,  \label{dder} \\
d^{\alpha } &=&\left( d^{i},d^{a}\right) \rightarrow \delta
^{\alpha }=\left( d^{i},\delta ^{a}=d^{a}+N_{i}^{b}d^{i}\right)
.  \label{ddif}
\end{eqnarray}%
The transforms can be considered as some particular case of frame
(vielbein) transforms of type
\begin{equation*}
\partial _{\alpha }\rightarrow \delta _{\alpha }=e_{\alpha }^{\beta
}\partial _{\beta }\mbox{ and }d^{\alpha }\rightarrow \delta
^{\alpha }=(e^{-1})_{\beta }^{\alpha }\delta ^{\beta },
\end{equation*}%
$e_{\alpha }^{\beta }(e^{-1})_{\beta }^{\gamma }=\delta _{\alpha
}^{\gamma }, $ when the ''tetradic'' coefficients $\delta
_{\alpha }^{\beta }$ are induced by using the Kronecker symbols
$\delta _{a}^{b},\delta _{j}^{i}$ and $N_{i}^{b}.$

The bases $\delta _{\alpha }$ and $\delta ^{\alpha }$ satisfy in
general some anholonomy conditions, for instance,
\begin{equation}
\delta _{\alpha }\delta _{\beta }-\delta _{\beta }\delta _{\alpha
}=W_{\alpha \beta }^{\gamma }\delta _{\gamma },  \label{anhol}
\end{equation}%
where $W_{\alpha \beta }^{\gamma }$ are called the anholonomy
coefficients.

Tensor fields on a vector bundle $\xi =\left( E,\mu ,M\right) $
provided with N--connection structure $N,$ we shall write $\xi
_{N},$ may be decomposed with in N--adapted form with respect to
the bases $\delta _{\alpha }$ and $\delta ^{\alpha },$ and their
tensor products. For instance, for a tensor of rang (1,1)
$T=\{T_{\alpha }^{~\beta }=\left(
T_{i}^{~j},T_{i}^{~a},T_{b}^{~j},T_{a}^{~b}\right) \}$ we have
\begin{equation}
T=T_{\alpha }^{~\beta }\delta ^{\alpha }\otimes \delta _{\beta
}=T_{i}^{~j}d^{i}\otimes \delta _{i}+T_{i}^{~a}d^{i}\otimes
\partial _{a}+T_{b}^{~j}\delta ^{b}\otimes \delta
_{j}+T_{a}^{~b}\delta ^{a}\otimes
\partial _{b}.  \label{dten}
\end{equation}

Every N--connection with coefficients $N_{i}^{b}$ $\
$automatically generates a linear connection on $\xi $ as $\Gamma
_{\alpha \beta }^{(N)\gamma }=\{N_{bi}^{a}=\partial
N_{i}^{a}(x,y)/\partial y^{b}\}$ which defines a covariant
derivative $D_{\alpha }^{(N)}A^{\beta }=\delta _{\alpha }A^{\beta
}+\Gamma _{\alpha \gamma }^{(N)\beta }A^{\gamma }.$

Another important characteristic of a N--connection is its
curvature $\Omega =\{\Omega _{ij}^{a}\}$ with the coefficients
\begin{equation*}
\Omega _{ij}^{a}=\delta _{j}N_{i}^{a}-\delta
_{i}N_{j}^{a}=\partial _{j}N_{i}^{a}-\partial
_{i}N_{j}^{a}+N_{i}^{b}N_{bj}^{a}-N_{j}^{b}N_{bi}^{a}.
\end{equation*}

In general, on a vector bundle we consider arbitrary linear
connection and, for instance, metric structure adapted to the
N--connection decomposition into vertical and horizontal
subbundles (one says that such objects are distinguished by the
N--connection, in brief, d--objects, like the d-tensor
(\ref{dten}), d--connection, d--metric:

\begin{itemize}
\item the coefficients of linear d--connections $\Gamma =\{\Gamma _{\alpha
\gamma }^{\beta }=\left(
L_{jk}^{i},L_{bk}^{a},C_{jc}^{i},C_{ac}^{b}\right) \}$ are
defined for an arbitrary covariant derivative $D$ on $\xi $ being
adapted to the $N$--connection structure as $D_{\delta _{\alpha
}}(\delta _{\beta })=\Gamma _{\beta \alpha }^{\gamma }\delta
_{\gamma }$ with the coefficients being invariant under
horizontal and vertical decomposition
\begin{equation*}
\quad D_{\delta _{i}}(\delta _{j})=L_{ji}^{k}\delta
_{k},~D_{\delta _{i}}(\partial _{a})=L_{ai}^{b}\partial
_{b},~D_{\partial _{c}}(\delta _{j})=C_{jc}^{k}\delta
_{k},~~D_{\partial _{c}}(\partial _{a})=C_{ac}^{b}\partial _{b}.
\end{equation*}

\item the d--metric structure $G=g_{\alpha \beta }\delta ^{a}\otimes \delta
^{b}$ which has the invariant decomposition as $g_{\alpha \beta
}=\left(
g_{ij},g_{ab}\right) $ following from%
\begin{equation}
G=g_{ij}(x,y)d^{i}\otimes d^{j}+g_{ab}(x,y)\delta ^{a}\otimes
\delta ^{b}. \label{dmetric}
\end{equation}
\end{itemize}

We may impose the condition that a d--metric and a d--connection
are compatible, i. e. there are satisfied the conditions
\begin{equation}
D_{\gamma }g_{\alpha \beta }=0.  \label{metrcond}
\end{equation}

With respect to the anholonomic frames (\ref{dder}) and
(\ref{ddif}), there is a linear connection, called the canonical
distinguished linear connection, which is similar to the metric
connection introduced by the Christoffel symbols in the case of
holonomic bases, i. e. being constructed
only from the metric components and satisfying the metricity conditions (\ref%
{metrcond}). It is parametrized by the coefficients,\ $\Gamma _{\
\beta \gamma }^{\alpha }=\left( L_{\ jk}^{i},L_{\ bk}^{a},C_{\
jc}^{i},C_{\ bc}^{a}\right) $ with the coefficients
\begin{eqnarray}
L_{\ jk}^{i} &=&\frac{1}{2}g^{in}\left( \delta _{k}g_{nj}+\delta
_{j}g_{nk}-\delta _{n}g_{jk}\right) ,  \label{dcon} \\
L_{\ bk}^{a} &=&\partial _{b}N_{k}^{a}+\frac{1}{2}h^{ac}\left(
\delta _{k}h_{bc}-h_{dc}\partial _{b}N_{k}^{d}-h_{db}\partial
_{c}N_{k}^{d}\right) ,
\notag \\
C_{\ jc}^{i} &=&\frac{1}{2}g^{ik}\partial _{c}g_{jk},\ C_{\ bc}^{a}=\frac{1}{%
2}h^{ad}\left( \partial _{c}h_{db}+\partial _{b}h_{dc}-\partial
_{d}h_{bc}\right) .  \notag
\end{eqnarray}%
We note that on Riemannian spaces the N--connection is an object
completely
defined by anholonomic frames, when the coefficients of frame transforms, $%
e_{\alpha }^{\beta }\left( u^{\gamma }\right) ,$ are parametrized
explicitly via certain values $\left( N_{i}^{a},\delta
_{i}^{j},\delta _{b}^{a}\right) , $ where $\delta _{i}^{j}$ $\
$and $\delta _{b}^{a}$ are the Kronecker symbols. By
straightforward calculations we can compute that the coefficients
of the Levi--Civita metric connection
\begin{equation*}
\Gamma _{\alpha \beta \gamma }^{\bigtriangledown }=g\left( \delta
_{\alpha },\bigtriangledown _{\gamma }\delta _{\beta }\right)
=g_{\alpha \tau }\Gamma _{\beta \gamma }^{\bigtriangledown \tau
},\,
\end{equation*}%
associated to a covariant derivative operator $\bigtriangledown ,$
satisfying the metricity condition $\bigtriangledown _{\gamma
}g_{\alpha \beta }=0$ for $g_{\alpha \beta }=\left(
g_{ij},h_{ab}\right) ,$
\begin{equation}
\Gamma _{\alpha \beta \gamma }^{\bigtriangledown
}=\frac{1}{2}\left[ \delta _{\beta }g_{\alpha \gamma }+\delta
_{\gamma }g_{\beta \alpha }-\delta _{\alpha }g_{\gamma \beta
}+g_{\alpha \tau }W_{\gamma \beta }^{\tau }+g_{\beta \tau
}W_{\alpha \gamma }^{\tau }-g_{\gamma \tau }W_{\beta \alpha
}^{\tau }\right] ,  \label{lcsym}
\end{equation}%
are given with respect to the anholonomic basis (\ref{ddif}) by
the coefficients
\begin{equation}
\Gamma _{\beta \gamma }^{\bigtriangledown \tau }=\left( L_{\
jk}^{i},L_{\ bk}^{a},C_{\ jc}^{i}+\frac{1}{2}g^{ik}\Omega
_{jk}^{a}h_{ca},C_{\ bc}^{a}\right) .  \label{lccon}
\end{equation}%
A specific property of off--diagonal metrics is that they can
define different classes of linear connections which satisfy the
metricity conditions for a given metric, or inversely, there is a
certain class of metrics which satisfy the metricity conditions
for a given linear connection. \ This result was originally
obtained by A. Kawaguchi \cite{kaw} (Details can be found in Ref.
\cite{ma}, see Theorems 5.4 and 5.5 in Chapter III, formulated
for vector bundles; here we note that similar proofs hold also on
manifolds enabled with anholonomic frames associated to a
N--connection structure).

With respect to anholonomic frames, we can not distinguish the
Levi--Civita connection as the unique both metric and torsionless
one. For instance, both linear connections (\ref{dcon}) and
(\ref{lccon}) contain anholonomically induced torsion
coefficients, are compatible with the same metric and transform
into the usual Levi--Civita coefficients for vanishing
N--connection and ''pure'' holonomic coordinates. This means that
to an off--diagonal metric in general relativity one may be
associated different covariant differential calculi, all being
compatible with the same metric structure (like in the
non--commutative geometry, which is not a surprising \ fact
because the anolonomic frames satisfy by definition some
non--commutative relations (\ref{anhol})). In such cases we have
to select a particular type of connection following some physical
or geometrical arguments, or to impose some conditions when there
is a single compatible linear connection constructed only from
the metric and N--coefficients. We
note that if $\Omega _{jk}^{a}=0$ the connections (\ref{dcon}) and (\ref%
{lccon}) coincide, i. e. $\Gamma _{\ \beta \gamma }^{\alpha
}=\Gamma _{\beta \gamma }^{\bigtriangledown \alpha }.$

\subsubsection{D--torsions and d--curvatures:}

The anholonomic coefficients $W_{\ \alpha \beta }^{\gamma }$ and
N--elongated derivatives give nontrivial coefficients for the
torsion tensor, $T(\delta _{\gamma },\delta _{\beta })=T_{\ \beta
\gamma }^{\alpha }\delta _{\alpha },$ where
\begin{equation}
T_{\ \beta \gamma }^{\alpha }=\Gamma _{\ \beta \gamma }^{\alpha
}-\Gamma _{\ \gamma \beta }^{\alpha }+w_{\ \beta \gamma }^{\alpha
},  \label{torsion}
\end{equation}%
and for the curvature tensor, $R(\delta _{\tau },\delta _{\gamma
})\delta _{\beta }=R_{\beta \ \gamma \tau }^{\ \alpha }\delta
_{\alpha },$ where
\begin{eqnarray}
R_{\beta \ \gamma \tau }^{\ \alpha } &=&\delta _{\tau }\Gamma _{\
\beta \gamma }^{\alpha }-\delta _{\gamma }\Gamma _{\ \beta \tau
}^{\alpha }  \notag
\\
&&+\Gamma _{\ \beta \gamma }^{\varphi }\Gamma _{\ \varphi \tau
}^{\alpha }-\Gamma _{\ \beta \tau }^{\varphi }\Gamma _{\ \varphi
\gamma }^{\alpha }+\Gamma _{\ \beta \varphi }^{\alpha }w_{\
\gamma \tau }^{\varphi }. \label{curvature}
\end{eqnarray}%
We emphasize that the torsion tensor on (pseudo) Riemannian
spacetimes is induced by anholonomic frames, whereas its
components vanish with respect to holonomic frames. All tensors
are distinguished (d) by the N--connection structure into
irreducible (horizontal--vertical) h--v--components, and are
called d--tensors. For instance, the torsion, d--tensor has the
following irreducible, nonvanishing, h--v--components,\ $T_{\
\beta \gamma }^{\alpha }=\{T_{\ jk}^{i},C_{\ ja}^{i},S_{\
bc}^{a},T_{\ ij}^{a},T_{\ bi}^{a}\},$ where
\begin{eqnarray}
T_{.jk}^{i} &=&T_{jk}^{i}=L_{jk}^{i}-L_{kj}^{i},\quad
T_{ja}^{i}=C_{.ja}^{i},\quad T_{aj}^{i}=-C_{ja}^{i},  \notag \\
T_{.ja}^{i} &=&0,\quad
T_{.bc}^{a}=S_{.bc}^{a}=C_{bc}^{a}-C_{cb}^{a},
\label{dtors} \\
T_{.ij}^{a} &=&-\Omega _{ij}^{a},\quad T_{.bi}^{a}=\partial
_{b}N_{i}^{a}-L_{.bi}^{a},\quad T_{.ib}^{a}=-T_{.bi}^{a}  \notag
\end{eqnarray}%
(the d--torsion is computed by substituting the
h--v--compo\-nents of the
canonical d--connection (\ref{dcon}) and anholonomy coefficients(\ref{anhol}%
) into the formula for the torsion coefficients (\ref{torsion})).

The curvature d-tensor has the following irreducible,
non-vanishing, h--v--compon\-ents\ $R_{\beta \ \gamma \tau }^{\
\alpha
}=%
\{R_{h.jk}^{.i},R_{b.jk}^{.a},P_{j.ka}^{.i},P_{b.ka}^{.c},S_{j.bc}^{.i},S_{b.cd}^{.a}\},
$\ where
\begin{eqnarray}
R_{h.jk}^{.i} &=&\delta _{k}L_{.hj}^{i}-\delta
_{j}L_{.hk}^{i}+L_{.hj}^{m}L_{mk}^{i}-L_{.hk}^{m}L_{mj}^{i}-C_{.ha}^{i}%
\Omega _{.jk}^{a},  \label{dcurvatures} \\
R_{b.jk}^{.a} &=&\delta _{k}L_{.bj}^{a}-\delta
_{j}L_{.bk}^{a}+L_{.bj}^{c}L_{.ck}^{a}-L_{.bk}^{c}L_{.cj}^{a}-C_{.bc}^{a}%
\Omega _{.jk}^{c},  \notag \\
P_{j.ka}^{.i} &=&\partial
_{a}L_{.jk}^{i}+C_{.jb}^{i}T_{.ka}^{b}-(\delta
_{k}C_{.ja}^{i}+L_{.lk}^{i}C_{.ja}^{l}-L_{.jk}^{l}C_{.la}^{i}-L_{.ak}^{c}C_{.jc}^{i}),
\notag \\
P_{b.ka}^{.c} &=&\partial
_{a}L_{.bk}^{c}+C_{.bd}^{c}T_{.ka}^{d}-(\delta
_{k}C_{.ba}^{c}+L_{.dk}^{c\
}C_{.ba}^{d}-L_{.bk}^{d}C_{.da}^{c}-L_{.ak}^{d}C_{.bd}^{c}),  \notag \\
S_{j.bc}^{.i} &=&\partial _{c}C_{.jb}^{i}-\partial
_{b}C_{.jc}^{i}+C_{.jb}^{h}C_{.hc}^{i}-C_{.jc}^{h}C_{hb}^{i},  \notag \\
S_{b.cd}^{.a} &=&\partial _{d}C_{.bc}^{a}-\partial
_{c}C_{.bd}^{a}+C_{.bc}^{e}C_{.ed}^{a}-C_{.bd}^{e}C_{.ec}^{a}
\notag
\end{eqnarray}
(the d--curvature components are computed in a similar fashion by
using the formula for curvature coefficients (\ref{curvature})).

\subsubsection{Einstein equations in d--variables}

In this subsection we write and analyze the Einstein equations on
spaces provided with anholonomic frame structures and associated
N--connections.

The Ricci tensor $R_{\beta \gamma }=R_{\beta ~\gamma \alpha
}^{~\alpha }$ has the d--components
\begin{eqnarray}
R_{ij} &=&R_{i.jk}^{.k},\quad R_{ia}=-^2P_{ia}=-P_{i.ka}^{.k},
\label{dricci} \\
R_{ai} &=&^1P_{ai}=P_{a.ib}^{.b},\quad R_{ab}=S_{a.bc}^{.c}.
\notag
\end{eqnarray}
In general, since $^1P_{ai}\neq ~^2P_{ia}$, the Ricci d-tensor is
non-symmetric (this could be with respect to anholonomic frames of
reference). The scalar curvature of the metric d--connection, $%
\overleftarrow{R}=g^{\beta \gamma }R_{\beta \gamma },$ is computed
\begin{equation}
{\overleftarrow{R}}=G^{\alpha \beta }R_{\alpha \beta
}=\widehat{R}+S, \label{dscalar}
\end{equation}
where $\widehat{R}=g^{ij}R_{ij}$ and $S=h^{ab}S_{ab}.$

By substituting (\ref{dricci}) and (\ref{dscalar}) into the
Einstein equations
\begin{equation}
R_{\alpha \beta }-\frac{1}{2}g_{\alpha \beta }R=\kappa \Upsilon
_{\alpha \beta },  \label{5einstein}
\end{equation}%
where $\kappa $ and $\Upsilon _{\alpha \beta }$ are respectively
the coupling constant and the energy--momentum tensor we obtain
the h-v-decomposition by N--connection of the Einstein equations
\begin{eqnarray}
R_{ij}-\frac{1}{2}\left( \widehat{R}+S\right) g_{ij} &=&\kappa
\Upsilon
_{ij},  \label{einsteq2} \\
S_{ab}-\frac{1}{2}\left( \widehat{R}+S\right) h_{ab} &=&\kappa
\Upsilon
_{ab},  \notag \\
^{1}P_{ai}=\kappa \Upsilon _{ai},\ ^{2}P_{ia} &=&\kappa \Upsilon
_{ia}. \notag
\end{eqnarray}%
The definition of matter sources with respect to anholonomic
frames is considered in Refs. \cite{vspinors,vmon1,ma}.

The vacuum 5D, locally anisotropic gravitational field equations,
in invariant h-- v--components, are written
\begin{eqnarray}
R_{ij} &=&0,S_{ab}=0,  \label{einsteq3} \\
^{1}P_{ai} &=&0,\ ^{2}P_{ia}=0.  \notag
\end{eqnarray}

We emphasize that vector bundles and even the (pseudo) Riemannian
space-times admit non--trivial torsion components, if
off--diagonal metrics and anholomomic frames are introduced into
consideration. This is a ''pure'' anholonomic frame effect: the
torsion vanishes for the Levi--Civita connection stated with
respect to a coordinate frame, but even this metric connection
contains some torsion coefficients if it is defined with respect
to anholonomic frames (this follows from the $W$--terms in
(\ref{lcsym})). For (pseudo) Riemannian spaces we conclude that
the Einstein theory transforms into an effective Einstein--Cartan
theory with anholonomically induced torsion if the general
relativity is formulated with respect to general frame bases
(both holonomic and anholonomic).

The N--connection geometry can be similarly formulated for a tangent bundle $%
TM$ of a manifold $M$ (which is used in Finsler and Lagrange geometry \cite%
{ma}), on cotangent bundle $T^{\ast }M$ and higher order bundles
(higher order Lagrange and Hamilton geometry \cite{miron}) as
well in the geometry of locally anisotropic superspaces
\cite{vsuper}, superstrings \cite{vstr2}, anisotropic spinor
\cite{vspinors} and gauge \ \cite{vgauge} theories or even on
(pseudo) Riemannian spaces provided with anholonomic frame
structures \cite{vmon2}.

\subsection{Nonlinear connections in projective modules}

{\quad }The nonlinear connection (N--connection) for
noncommutative spaces can be defined similarly to commutative
spaces by considering instead of usual vector bundles theirs
noncommutative analogs defined as finite projective modules over
noncommutative algebras. The explicit constructions depend on the
type of differential calculus we use for definition of tangent
structures and theirs maps.

In general, there can be several differential calculi over a given algebra $%
\mathcal{A}$ (for a more detailed discussion within the context of
noncommutative geometry see Refs. \cite{connes,madore,dubois}; a
recent approach is connected with Lie superalgebra structures on
the space of multiderivations \cite{giun}). \ For simplicity, in
this work we fix a differential calculus on $\mathcal{A},$ which
means that we choose a
(graded) algebra $\Omega ^{\ast }(\mathcal{A})=\cup _{p}\Omega ^{p}(\mathcal{%
A})$ which gives a differential structure to $\mathcal{A}.$ The elements of $%
\Omega ^{p}(\mathcal{A})$ are called $p$--forms. There is a
linear map $d$ which takes $p$--forms into $(p+1)$--forms and
which satisfies a graded
Leibniz rule as well the condition $d^{2}=0.$ By definition $\Omega ^{0}(%
\mathcal{A})=\mathcal{A}.$

The differential $df$ of a real or complex variable on a vector
bundle $\xi _{N}$
\begin{eqnarray*}
df &=&\delta _{i}f~dx^{i}+\partial _{a}f~\delta y^{a}, \\
\delta _{i}f~ &=&\partial _{i}f-N_{i}^{a}\partial _{a}f~,~\delta
y^{a}=dy^{a}+N_{i}^{a}dx^{i}
\end{eqnarray*}%
in the noncommutative case is replaced by a distinguished
commutator
(d--commutator)%
\begin{equation*}
\overline{d}f=\left[ F,f\right] =\left[ F^{[h]},f\right] +\left[ F^{[v]},f%
\right]
\end{equation*}%
where the operator $F^{[h]}$ $\ (F^{[v]})$ is acting on the
horizontal (vertical) projective submodule being defined by some
fixed differential
calculus $\Omega ^{\ast }(\mathcal{A}^{[h]})$ ($\Omega ^{\ast }(\mathcal{A}%
^{[v]}))$ on the so--called horizontal (vertical) $\mathcal{A}^{[h]}$ ($%
\mathcal{A}^{[v]})$ algebras.

Let us consider instead of a vector bundle $\xi $ $\ $an $\mathcal{A}$%
--module $\mathcal{E}$ being projective and of finite type. For a
fixed
differential calculus on $\mathcal{E}$ we define the tangent structures $T%
\mathcal{E}$ \ and $TM.\,$ A nonlinear connection$\,N$ in an $\mathcal{A}$%
--module $\mathcal{E}$ is defined by an exact sequence of finite projective $%
\mathcal{A}$--moduli
\begin{equation*}
0\rightarrow V\mathcal{E}\rightarrow
T\mathcal{E}/V\mathcal{E}\rightarrow 0,
\end{equation*}%
where all subspaces are constructed as in the commutative case
with that difference that the vector bundle objects are
substituted by theirs projective modules equivalents. A
projective module provided with N--connection structures will be
denoted as $\mathcal{E}_{N}.$ All objects on a $\mathcal{E}_{N}$
\ have a distinguished invariant character with respect to the
horizontal and vertical subspaces.

To understand how the N--connection structure may be taken into
account on noncommutative spaces we analyze in the next
subsection an example.

\subsection{Commutative and noncommutative gauge d--fields}

Let us consider a vector bundle $\xi _{N}$ and a another vector bundle $%
\beta =\left( B,\pi ,\xi _{N}\right) $ with $\pi :B\rightarrow
\xi _{N}$ with a typical $k$-dimensional vector fiber. In local
coordinates a linear connection (a gauge field) in $\beta $ is
given by a collection of
differential operators%
\begin{equation*}
\bigtriangledown _{\alpha }=D_{\alpha }+B_{\alpha }(u),
\end{equation*}%
acting on $T\xi _{N}$ where
\begin{equation*}
D_{\alpha }=\delta _{\alpha }\pm \Gamma _{\cdot \alpha }^{\cdot }%
\mbox{ with }D_{i}=\delta _{i}\pm \Gamma _{\cdot i}^{\cdot
}\mbox{ and }D_{a}=\partial _{a}\pm \Gamma _{\cdot a}^{\cdot }
\end{equation*}
is a d--connection in $\xi _{N}$ ($\alpha =1,2,...,n+m),$ with
$\delta
_{\alpha }$ N--elongated as in (\ref{dder}), $u=(x,y)\in \xi _{N}$ and $%
B_{\alpha }$ are $k\times k$--matrix valued functions. For every
vector field
\begin{equation*}
X=X^{\alpha }(u)\delta _{\alpha }=X^{i}(u)\delta
_{i}+X^{a}(u)\partial _{a}\in T\xi _{N}
\end{equation*}
we can consider the operator
\begin{equation}
X^{\alpha }(u)\bigtriangledown _{\alpha }(f\cdot s)=f\cdot
\bigtriangledown _{X}s+\delta _{X}f\cdot s  \label{rul1c}
\end{equation}%
for any section $s\in \mathcal{B}$ \ and function $f\in C^{\infty
}(\xi
_{N}),$ where%
\begin{equation*}
\delta _{X}f=X^{\alpha }\delta _{\alpha }~\mbox{ and
}\bigtriangledown _{fX}=f\bigtriangledown _{X}.
\end{equation*}%
In the simplest definition we assume that there is a Lie algebra $\mathcal{GL%
}B$ that acts on associative algebra $B$ by means of infinitesimal
automorphisms (derivations). This means that we have linear operators $%
\delta _{X}:B\rightarrow B$ which linearly depend on $X$ and satisfy%
\begin{equation*}
\delta _{X}(a\cdot b)=(\delta _{X}a)\cdot b+a\cdot (\delta _{X}b)
\end{equation*}%
for any $a,b\in B.$ The mapping $X\rightarrow \delta _{X}$ is a
Lie algebra homomorphism, i. e. $\delta _{\lbrack X,Y]}=[\delta
_{X},\delta _{Y}].$

Now we consider respectively instead of vector bundles $\xi $ and
$\beta $ the finite \ projective $\mathcal{A}$--module
$\mathcal{E}_{N},$ provided
with N--connection structure, and the finite projective $\mathcal{B}$%
--module $\mathcal{E}_{\beta }.$

A d--connection $\bigtriangledown _{X\text{ }}$on
$\mathcal{E}_{\beta }$ is by definition a set of linear
d--operators, adapted to the N--connection
structure, depending linearly on $X$ and satisfying the Leibniz rule%
\begin{equation}
\bigtriangledown _{X}(b\cdot e)=b\cdot \bigtriangledown
_{X}(e)+\delta _{X}b\cdot e  \label{rul1n}
\end{equation}%
for any $e\in \mathcal{E}_{\beta }$ and $b\in \mathcal{B}.$ The rule (\ref%
{rul1n}) is a noncommutative generalization of (\ref{rul1c}). We
emphasize that both operators $\bigtriangledown _{X}$ and $\delta
_{X}$ are distinguished by the N--connection structure and that
the difference of two such linear d--operators, $\bigtriangledown
_{X}-\bigtriangledown _{X}^{\prime }$ commutes with action of $B$
on $\mathcal{E}_{\beta },$ which is an endomorphism of
$\mathcal{E}_{\beta }.$ Hence, if we fix some fiducial
connection $\bigtriangledown _{X}^{\prime }$ (for instance, $%
\bigtriangledown _{X}^{\prime }=D_{X})$ on $\mathcal{E}_{\beta }$
an arbitrary connection has the form
\begin{equation*}
\bigtriangledown _{X}=D_{X}+B_{X},
\end{equation*}%
where $B_{X}\in End_{B}\mathcal{E}_{\beta }$ depend linearly on
$X.$

The curvature of connection $\bigtriangledown _{X}$ is a
two--form $F_{XY}$ which values linear operator in $\mathcal{B}$
and measures a deviation of mapping $X\rightarrow
\bigtriangledown _{X}$ from being a Lie algebra
homomorphism,%
\begin{equation*}
F_{XY}=[\bigtriangledown _{X},\bigtriangledown
_{Y}]-\bigtriangledown _{\lbrack X,Y]}.
\end{equation*}%
The usual curvature d--tensor is defined as
\begin{equation*}
F_{\alpha \beta }=\left[ \bigtriangledown _{\alpha
},\bigtriangledown _{\beta }\right] -\bigtriangledown _{\lbrack
\alpha ,\beta ]}.
\end{equation*}

The simplest connection on a finite projective $\mathcal{B}$--module $%
\mathcal{E}_{\beta }$ is to be specified by a projector $P:\mathcal{B}%
^{k}\otimes \mathcal{B}^{k}$ when the d--operator $\delta _{X}$
acts
naturally on the free module $\mathcal{B}^{k}.$ The operator $%
\bigtriangledown _{X}^{LC}=P\cdot \delta _{X}\cdot P$ $\ $\ is
called the Levi--Civita operator and satisfy the condition
$Tr[\bigtriangledown _{X}^{LC},\phi ]=0$ for any endomorphism
$\phi \in End_{B}\mathcal{E}_{\beta }.$ From this identity, and
from the fact that any two connections differ by an endomorphism
that
\begin{equation*}
Tr[\bigtriangledown _{X},\phi ]=0
\end{equation*}%
for an arbitrary connection $\bigtriangledown _{X}$ and an
arbitrary endomorphism $\phi ,$ that instead of $\bigtriangledown
_{X}^{LC}$ we may consider equivalently the canonical
d--connection, constructed only from d-metric and N--connection
coefficients.

\section{ Distinguished Spectral Triples}

In this section we develop the basic ingredients introduced by A. Connes %
\cite{connes} to define the analogue of differential calculus for
noncommutative distinguished algebras. The N--connection
structures distinguish a commutative or a noncommutative spaces
into horizontal and vertical subspaces. The geometric objects
possess a distinguished invariant character with respect to a
such splitting. The basic idea in definition of spectral triples
generating locally anisotropic spaces (Rimannian spaces with
anholonomic structure, or, for more general constructions,
Finsler and Lagrange spaces) is to consider pairs of noncommutative algebras $\mathcal{A}%
_{[d]}=(\mathcal{A}_{[h]},\mathcal{A}_{[v]}),$ given by
respective pairs of elements $a=\left( a_{[h]},a_{[v]}\right) \in
\mathcal{A}_{[d]},$ called also distinguished algebras (in brief,
d--algebras), together with d-operators
$D_{[d]}=(D_{[h]},D_{[v]})$ on a Hilbert space $\mathcal{H}$ (for
simplicity we shall consider one Hilbert space, but a more general
construction can be provided for Hilbert d-spaces,
$\mathcal{H}_{[d]}=\left(
\mathcal{H}_{[h]},\mathcal{H}_{[v]}\right) .$

The formula of Wodzicki--Adler--Manin--Guillemin residue (see, for instance, %
\cite{landi}) may be writen for vector bundles provided with
N--connection
structure. It is necessary to introduce the N--elongated differentials (\ref%
{dder}) in definition of the measure: Let $Q$ be a
pseudo--differential
poprator of order $-n$ acting on sections of a complex vector bundle $%
E\rightarrow M$ over an $n$--dimensional compact Riemannian
manifold $M.$ The residue $ResQ$ of $Q$ is defined by the formula
\begin{equation*}
{Re}sQ=:\frac{1}{n\left( 2\pi \right) ^{n}}\int\limits_{S^{\ast
}M}tr_{E}\sigma _{-n}(Q)\delta \mu ,
\end{equation*}%
where $\sigma _{-n}(Q)$ is the principal symbol (a matrix--valued
function on $T^{\ast }M$ which is homogeneous of degree $-n$ in
the fiber coordinates), \ the integral is taken over the unit
co--sphere $S^{\ast }M=\{(x,y)\in T^{\ast }M:||y||=1\}\subset
T^{\ast }M,$ the $tr_{E}$ is the matrix trace over ''internal
indices'' and the measures $\delta \mu =dx^{i}\delta y^{a}.$

A spectral d--triple $\left[
\mathcal{A}_{[d]},\mathcal{H},D_{[d]}\right] $ is given by an
involutive d--algebra of d--operators $D^{[d]}$ consisting from
pairs of bounded operators $D_{[h]}$ and $D_{[v]}$ $\ $on the
Hilbert
space $\mathcal{H},$ together with the self--adjoint operation $%
D_{[d]}=D_{[d]}^{\ast }$ for respective $h$- and $v$--components on $%
\mathcal{H}$ being satisfied the properties:

\begin{enumerate}
\item The resolvents $(D_{[h]}-\lambda _{\lbrack h]})^{-1}$ and $%
(D_{[v]}-\lambda _{\lbrack v]})^{-1},$ $\lambda _{\lbrack
h]},\lambda _{\lbrack v]}\in \R,$ are compact operators on
$\mathcal{H};$

\item The commutators $\left[ D_{[h,]},a_{[h]}\right] \doteqdot
D_{[h]}a_{[h]}-a_{[h]}D_{[h]}\in \mathcal{B}(\mathcal{H})$ and
$\left[
D_{[v,]},a_{[v]}\right] \doteqdot D_{[v]}a_{[v]}-a_{[v]}D_{[v]}\in \mathcal{B%
}(\mathcal{H})$ for any $a\in \mathcal{A}_{[d]},$ where by $\mathcal{B}(%
\mathcal{H})$ we denote the algebra of bounded operators on
$\mathcal{H}.$
\end{enumerate}

The $h(v)$--component of a d--triple is said to be even if there is a ${\Z}%
_{2}$--grading for $\mathcal{H},$ i. e. an operator $\Upsilon $ on $\mathcal{%
H}$ such that
\begin{equation*}
\Upsilon =\Upsilon ^{\ast },\Upsilon ^{2}=1,~\Upsilon
D_{[h(v)]}-D_{[h(v)]}\Upsilon =0,~\Upsilon a-a\Upsilon =0
\end{equation*}%
for every $a\in \mathcal{A}_{[d]}.$ If such a grading does not exist, the $%
h(v)$--component of a d--triple is said to be odd.

\subsection{Canonical triples over vector bundles}

The basic examples of spectral triples in connections with
noncommutative field theory and geometry models were constructed
by means of the Dirac operator on a closed $n$--dimensional
Riemannian spin manifold $\left( M,g\right) $ \cite{connes,cl}. \
In order to generate by using functional methods some anisotropic
geometries, it is necessary to generalize the approach to vector
and covector bundles provided with compatible N--connection,
d--connection and metric structures. The theory of spinors on
locally anisotropic spaces was developed in Refs.
\cite{vspinors,vmon2}. \ This section is devoted to the spectral
d--triples defined by the Dirac operators on closed regions of
$\left( n+m\right) $--dimensional spin--vector manifolds. We note
that if we deal with off--diagonal metrics and/or anholonomic
frames there is an infinite number of d--connections which are
compatible with d--metric and N--connection structures, see
discussion and details in Ref. \cite{vexsol}. For simplicity, we
restrict
our consideration only to the Euclidean signature of metrics of type (\ref%
{dmetric}) (on attempts to define triples with Minkowskian
signatures see, for instance, Refs. \cite{hawkins}).

For a spectral d--triple $\left[
\mathcal{A}_{[d]},\mathcal{H},D_{[d]}\right] $ associated to a
vector bundle $\xi _{N}$ one takes the components:

\begin{enumerate}
\item $\mathcal{A}_{[d]}=\mathcal{F}(\xi _{N})$ is the algebra of complex
valued functions on $\xi _{N}.$

\item $\mathcal{H}=L^{2}(\xi _{N},S)$ is the Hilbert space of square
integrable sections of the irreducible d--spinor bundle (of rank $%
2^{(n+m)/2} $ over $\xi _{N}$ \cite{vspinors,vmon2}. The scalar product in $%
L^{2}(\xi _{N},S)$ is the defined by the measure associated to
the d--metric
(\ref{dmetric}),%
\begin{equation*}
(\psi ,\phi )=\int \delta \mu (g)\overline{\psi }(u)\phi (u)
\end{equation*}%
were the bar indicates to the complex conjugation and the scalar
product in d--spinor space is the natural one in
$\C^{2[n/2]}\oplus \C^{2[m/2]}.$

\item $D$ is a Dirac d--operator associated to one of the d--metric
compatible d--connecti\-on, for instance, with the Levi--Civita
connection,
canonical d--connection or another one, denoted with a general symbol $%
\Gamma =\Gamma _{\mu }\delta u^{\mu }.$
\end{enumerate}

We note that the elements of the algebra $\mathcal{A}_{[d]}$ acts
as
multiplicative operators on $\mathcal{H},$%
\begin{equation*}
(a\psi )(u)=:f(u)\psi (u),
\end{equation*}%
for every $a\in \mathcal{A}_{[d]},\psi \in \mathcal{H}.$

\subsubsection{Distinguished spinor structures \label{spinorsubsection}}

Let us analyze the connection between d--spinor structures and
spectral d--triples over a vector bundle $\xi _{N}.$ One consider
a $(n+m)$--bein (frame) decomposition of the d--metric $g_{\alpha
\beta }$ (\ref{dmetric})
(and its inverse $g^{\alpha \beta }),$%
\begin{equation*}
g^{\alpha \beta }(u)=e_{\underline{\alpha }}^{\alpha
}(u)e_{\underline{\beta
}}^{\beta }(u)\eta ^{\underline{\alpha }\underline{\beta }},~\eta _{%
\underline{\alpha }\underline{\beta }}=e_{\underline{\alpha
}}^{\alpha }(u)e_{\underline{\beta }}^{\beta }(u)g_{\alpha \beta
},
\end{equation*}%
$\eta _{\underline{\alpha }\underline{\beta }}$ it the diagonal Euclidean $%
(n+m)$--metric, which is adapted to the N--connection structure
because the coefficients $g_{\alpha \beta }$ are defined with
respect to the dual N--distinguish\-ed basis (\ref{ddif}). We can
define compatible with this
decomposition d--connections $\Gamma _{\underline{\beta }\mu }^{\underline{%
\alpha }}$ (for instance, the Levi--Civita connection, which with
respect to anholonomic frames contains torsions components, or
the canonical d--connection), defined by
\begin{equation*}
D_{\mu }e_{\underline{\beta }}=\Gamma _{\underline{\beta }\mu }^{\underline{%
\alpha }}e_{\underline{\alpha }},
\end{equation*}%
as the solution of the equations%
\begin{equation*}
\delta _{\mu }e_{v}^{\underline{\nu }}-\delta _{\nu }e_{\mu }^{\underline{%
\nu }}=\Gamma _{\underline{\beta }\mu }^{\underline{\nu }}e_{v}^{\underline{%
\beta }}-\Gamma _{\underline{\beta }\nu }^{\underline{\nu }}e_{\mu }^{%
\underline{\beta }}.
\end{equation*}

We define by $C\left( \xi _{N}\right) $ the Clifford bundle over
\ $\xi _{N}$ with the fiber at $u\in \xi _{N}$ being just the
complexified Clifford d--algebra $Cliff_{\C}\left( T_{u}^{\ast
}\xi _{N}\right) ,$ $T_{u}^{\ast }\xi _{N}$ being dual to
$T_{u}\xi _{N},$ and $\Gamma \lbrack \xi _{N},C\left( \xi
_{N}\right) ]$ is the module of corresponding sections. By
defining the maps
\begin{equation*}
\gamma \left( \delta ^{\alpha }\right) =\left( \gamma \left(
d^{i}\right) ,\gamma \left( \delta ^{a}\right) \right) \doteqdot
\gamma ^{\alpha }(u)=\gamma ^{\underline{\alpha
}}e_{\underline{\alpha }}^{\alpha
}(u)=\left( \gamma ^{\underline{\alpha }}e_{\underline{\alpha }%
}^{i}(u),\gamma ^{\underline{\alpha }}e_{\underline{\alpha
}}^{a}(u)\right) ,
\end{equation*}%
extended as an algebra map by $\mathcal{A}_{[d]}$--linearity, we
construct
an algebra morphism%
\begin{equation}
\gamma :\Gamma \left( \xi _{N},C\left( \xi _{N}\right) \right)
\rightarrow \mathcal{B}(\mathcal{H}).  \label{gammamorph}
\end{equation}%
The indices of the ''curved'' $\gamma ^{\alpha }(u)$ and ''flat'' $\gamma ^{%
\underline{\alpha }}$ gamma matrices can be lowered by using
respectively
the d-metric components $g_{\alpha \beta }$ $(u)$ and $\eta _{\underline{%
\alpha }\underline{\beta }},$ i. e. $\gamma _{\beta }(u)$
=$\gamma ^{\alpha
}(u)$ $g_{\alpha \beta }$ $(u)$ and $\ \gamma _{\underline{\beta }}=\gamma ^{%
\underline{\alpha }}\eta _{\underline{\alpha }\underline{\beta
}}.$ We take the gamma matrices to be Hermitian and to obey the
relations,
\begin{eqnarray*}
\gamma _{\alpha }\gamma _{\beta }+\gamma _{\beta }\gamma _{\alpha
} &=&-2g_{\alpha \beta }~\left( \gamma _{i}\gamma _{j}+\gamma
_{j}\gamma _{i}=-2g_{ij},~\gamma _{a}\gamma _{b}+\gamma
_{b}\gamma _{a}=-2g_{ab}\right)
, \\
\gamma _{\underline{\alpha }}\gamma _{\underline{\beta }}+\gamma _{%
\underline{\beta }}\gamma _{\underline{\alpha }} &=&-2\eta _{\underline{%
\alpha }\underline{\beta }}.
\end{eqnarray*}%
Every d--connection $\Gamma _{\underline{\beta }\mu
}^{\underline{\nu }}$ can be shifted as a d--covariant operator
$\bigtriangledown _{\mu }^{[S]}=\left( \bigtriangledown
_{i}^{[S]},\bigtriangledown
_{a}^{[S]}\right) $ on the bundle of d--spinors,%
\begin{equation*}
\bigtriangledown _{\mu }^{[S]}=\delta _{\mu }+\frac{1}{2}\Gamma _{\underline{%
\alpha }\underline{\beta }\mu }\gamma ^{\underline{\alpha }}\gamma ^{%
\underline{\beta }},~\Gamma _{\mu }^{[S]}=\frac{1}{2}\Gamma _{\underline{%
\alpha }\underline{\beta }\mu }\gamma ^{\underline{\alpha }}\gamma ^{%
\underline{\beta }},
\end{equation*}%
which defines the Dirac d--operator
\begin{equation}
_{\lbrack d]}D=:\gamma \circ \bigtriangledown _{\mu
}^{[S]}=\gamma ^{\alpha
}(u)\left( \delta _{\mu }+\Gamma _{\mu }^{[S]}\right) =\gamma ^{\underline{%
\alpha }}e_{\underline{\alpha }}^{\mu }\left( \delta _{\mu
}+\Gamma _{\mu }^{[S]}\right) .  \label{dirac}
\end{equation}%
Such formulas were introduced in Refs. \cite{vspinors} for
distinguished spinor bundles (of first and higher order). In this
paper we revise them in connection to spectral d--triples and
noncommutative geometry. \ On such spaces one also holds a
variant of Lichnerowicz formula \cite{berline}\ for
the square of the Dirac d--operator%
\begin{equation}
_{\lbrack d]}D^{2}=\bigtriangledown ^{\lbrack S]}+\frac{1}{4}{\overleftarrow{%
R}},  \label{lich}
\end{equation}%
where the formulas for the scalar curvature ${\overleftarrow{R}}$
is given in (\ref{dscalar}) and
\begin{equation*}
\bigtriangledown ^{\lbrack S]}=-g^{\mu \nu }\left(
\bigtriangledown _{\mu }^{[S]}\bigtriangledown _{v}^{[S]}-\Gamma
_{\mu \nu }^{\rho }\bigtriangledown _{\rho }^{[S]}\right) .
\end{equation*}

In a similar manner as in Ref. \cite{landi} but reconsidering all
computations on a vector bundle $\xi _{N}$ we can prove that for
every d--triple $\left[
\mathcal{A}_{[d]},\mathcal{H},D_{[d]}\right] $ one holds the
properties:

\begin{enumerate}
\item The vector bundle $\xi _{N}$ is the structure space of the algebra $%
\overline{\mathcal{A}}_{[d]}$ of continuous functions on $\xi
_{N}$ (the bar here points to the norm closure of
$\mathcal{A}_{[d]}).$

\item The geodesic distance $\rho $ between two points $p_{1},p_{2}\in $ $%
\xi _{N}$ is defined by using the Dirac d--operator,%
\begin{equation*}
\rho \left( p_{1},p_{2}\right) =\sup_{f\in }\left\{
|f(p)-f(q)|:~||[D_{[d]},f]||\leq 1\right\} .
\end{equation*}

\item The Dirac d--operator also defines the Riemannian measure on $\xi
_{N}, $%
\begin{equation*}
\int_{\xi _{N}}f=c\left( n+m\right) tr_{\Gamma }\left(
f|D_{[d]}|^{-(n+m)}\right)
\end{equation*}%
for every $f\in \mathcal{A}_{[d]}$ and $c\left( n+m\right)
=2^{[n+m-(n+m)/2-1]}\pi ^{(n+m)/2}(n+m)\Gamma \left(
\frac{n+m}{2}\right) ,\Gamma $ being the gamma function.
\end{enumerate}

The spectral d--triple \ formalism has the same properties as the
usual one with that difference that we are working on spaces
provided with N--connection structures and the bulk of
constructions and objects are distinguished by this structure.

\subsubsection{Noncommutative differential forms}

To construct a differential algebra of forms out a spectral
d--triple $\left[ \mathcal{A}_{[d]},\mathcal{H},D_{[d]}\right] $
one follows universal graded differential d--algebras \ defined
as couples of universal ones, respectively associated to the
$h$-- and $v$--components of some splitting to subspaces defined
by N--connection structures. Let $\mathcal{A}_{[d]}$ be an
associative d--algebra (for simplicity, with unit) over the field
of complex numbers $\C.$ The universal d--algebra of differential
forms $\Omega \mathcal{A}_{[d]}=\oplus _{p}\Omega
^{p}\mathcal{A}_{[d]}$ is introduced as a graded d--algebra when
$\Omega ^{0}\mathcal{A}_{[d]}=\mathcal{A}_{[d]}$ and the space
$\Omega ^{1}\mathcal{A}_{[d]}$ of one--forms is generated as a
left $\mathcal{A}_{[d]}$--module by symbols of degree $\delta
a,a\in
\mathcal{A}_{[d]}$ satisfying the properties%
\begin{equation*}
\delta (ab)=(\delta a)b+a\delta b\mbox{ and }\delta (\alpha
a+\beta b)=\alpha (\delta a)+\beta \delta b
\end{equation*}%
from which follows $\delta 1=0$ which in turn implies $\delta
\C=0.$ These relations state the Leibniz rule for the map
\begin{equation*}
\delta :\mathcal{A}_{[d]}\rightarrow \Omega ^{1}\mathcal{A}_{[d]}
\end{equation*}%
An element $\varpi \in \Omega ^{1}\mathcal{A}_{[d]}$ is expressed
as a
finite sum of the form%
\begin{equation*}
\varpi
=\sum\limits_{\underline{i}}a_{\underline{i}}b_{\underline{i}}
\end{equation*}%
for $a_{\underline{i}},b_{\underline{i}}\in \mathcal{A}_{[d]}.$ The left $%
\mathcal{A}_{[d]}$--module $\Omega ^{1}\mathcal{A}_{[d]}$ can be
also endowed with a structure of right
$\mathcal{A}_{[d]}$--module if the
elements are imposed to satisfy the conditions%
\begin{equation*}
\left( \sum\limits_{\underline{i}}a_{\underline{i}}\delta b_{\underline{i}%
}\right) c=:\sum\limits_{\underline{i}}a_{\underline{i}}(\delta b_{%
\underline{i}})c=\sum\limits_{\underline{i}}a_{\underline{i}}\delta (b_{%
\underline{i}}c)-\sum\limits_{\underline{i}}a_{\underline{i}}b_{\underline{i}%
}\delta c.
\end{equation*}

Given a spectral d--triple $\left[ \mathcal{A}_{[d]},\mathcal{H},D_{[d]}%
\right] ,$ one constructs and exterior d--algebra of forms by
means of a suitable representation of the universal algebra
$\Omega \mathcal{A}_{[d]}$ in the d--algebra of bounded operators
on $\mathcal{H}$ by considering the
map%
\begin{eqnarray*}
\pi &:&\Omega \mathcal{A}_{[d]}\rightarrow \mathcal{B}(\mathcal{H}), \\
\pi \left( a_{0}\delta a_{1}...\delta a_{p}\right) &=&:a_{0}\left[ D,a_{1}%
\right] ...\left[ D,a_{p}\right]
\end{eqnarray*}%
which is a homomorphism since both $\delta $ and $\left[
D,.\right] $ are
distinguished derivations on $\mathcal{A}_{[d]}.$ More than that, since $%
\left[ D,a\right] ^{\ast }=-\left[ D,a^{\ast }\right] ,$ we have
$\pi \left(
\varpi \right) ^{\ast }=\pi \left( \varpi ^{\ast }\right) $ for any d--form $%
\varpi \in \Omega \mathcal{A}_{[d]}$ and $\pi $ being a $\ast $%
--homomorphism.

Let $J_{0}=:\oplus _{p}J_{0\text{\ }}^{p}$ be the graded
two--sided ideal of $\Omega \mathcal{A}_{[d]}$ given by
$J_{0\text{\ }}^{p}=:\{\pi \left( \varpi \right) =0\}$ when
$J=J_{0}+\delta J_{0}$ is a graded differential two--sided ideal
of $\Omega \mathcal{A}_{[d]}.$ At the next step we can
define the graded differential algebra of Connes' forms over the d--algebra $%
\mathcal{A}_{[d]}$ as%
\begin{equation*}
\Omega _{D}\mathcal{A}_{[d]}=:\Omega \mathcal{A}_{[d]}/J\simeq
\pi \left( \Omega \mathcal{A}_{[d]}\right) /\pi \left( \delta
J_{0}\right) .
\end{equation*}%
It is naturally graded by the degrees of $\Omega
\mathcal{A}_{[d]}$ and $J$
with the space of $p$--forms being given by $\Omega _{D}^{p}\mathcal{A}%
_{[d]}=\Omega ^{p}\mathcal{A}_{[d]}/J^{p}.$ Being $J$ a
differential ideal,
the exterior differential $\delta $ defines a differential on $\Omega _{D}%
\mathcal{A}_{[d]},$%
\begin{equation*}
\delta :\Omega _{D}^{p}\mathcal{A}_{[d]}\rightarrow \Omega _{D}^{p+1}%
\mathcal{A}_{[d]},~\delta \lbrack \varpi ]=:\left[ \delta \varpi
\right]
\end{equation*}%
with $\varpi \in \Omega _{D}^{p}\mathcal{A}_{[d]}$ and $[\varpi
]$ being the corresponding class in $\Omega
_{D}^{p}\mathcal{A}_{[d]}.$

We conclude that the theory of distinguished d--forms generated by
d--algebras, as well of the graded differential d--algebra of
Connes' forms, is constructed in a usual form (see Refs.
\cite{connes,landi}) but for two subspaces (the horizontal and
vertical ones) defined by a N--connection structure.

\subsubsection{The exterior d--algebra}

The differential d--form formalism when applied to the canonical d--triple $%
\left[ \mathcal{A}_{[d]},\mathcal{H},D_{[d]}\right] $ over an
ordinary vector bundle $\xi _{N}$ provided with N--connection
structure reproduce the usual exterior d--aplgebra over this
vector bundle. Consider our d--triple
on a closed $\left( n+m\right) $--dimensional Riemannian $spin^{c}\,$%
manifold as described in subsection \ref{spinorsubsection} when $\mathcal{A}%
_{[d]}=\mathcal{F}(\xi _{N})$ is the algebra of smooth complex
valued functions on $\xi _{N}$ and $\mathcal{H}=L^{2}(\xi
_{N},S)$ is the Hilbert space of square integrable sections of
the irreducible d--spinor bundle (of
rank $2^{(n+m)/2}$ over $\xi _{N}.$ We can identify%
\begin{equation}
\pi \left( \delta f\right) =:[_{[d]}D,f]=\gamma ^{\mu }(u)\delta
_{\mu }f=\gamma \left( \delta _{\xi _{N}}f\right)  \label{pi1}
\end{equation}%
for every $f\in \mathcal{A}_{[d]},$ see the formula for the Dirac
d--operator (\ref{dirac}), \ where $\gamma $ is the d--algebra morphism \ (%
\ref{gammamorph}) and $\delta _{\xi _{N}}$ denotes the usual
exterior
derivative on $\xi _{N}.$ In a more general case, with $f_{[i]}\in \mathcal{A%
}_{[d]},[i]=[1],...,[p],$ we can write
\begin{equation}
\pi \left( f_{[0]}\delta f_{[1]}...\delta f_{[p]}\right)
=:f_{[0]}[_{[d]}D,f_{[1]}]...[_{[d]}D,f_{[p]}]=\gamma \left(
f_{[0]}\delta _{\xi _{N}}f_{[1]}\cdot ...\cdot \delta _{\xi
_{N}}f_{[p]}\right) , \label{pi2}
\end{equation}%
where the d--differentials $\delta _{\xi _{N}}f_{[1]}$ are
regarded as sections of the Clifford d--bundle $C_{1}(\xi _{N}),$
while $f_{[i]}$ can be thought of as sections of $C_{0}(\xi
_{N})$ and the $dot$ $\cdot $ the Clifford product in the fibers
of $C(\xi _{N})=\oplus _{k}C_{k}(\xi _{N}),$ see details in Refs.
\cite{vspinors,vmon2}.

A generic differential 1--form on $\xi _{N}$ can be written as $%
\sum_{[i]}f_{0}^{[i]}\delta _{\xi _{N}}f_{1}^{[i]}$ with $%
f_{0}^{[i]},f_{1}^{[i]}$ $\in \mathcal{A}_{[d]}.$ Following the definitions (%
\ref{pi1}) and (\ref{pi2}), we can identify the distinguished
Connes' 1--forms $\Omega _{D}^{p}\mathcal{A}_{[d]}$ with the
usual distinguished differential 1--forms, i. e.
\begin{equation*}
\Omega _{D}^{p}\mathcal{A}_{[d]}\simeq \Lambda ^{p}\left( \xi
_{N}\right) .
\end{equation*}

For each $u\in \xi _{N},$ we can introduce a natural filtration
for the Clifford d--algebra, $C_{u}(\xi _{N})=\cup C_{u}^{(p)},$
where $C_{u}^{(p)}$ is spanned by products of type $\chi
_{\lbrack 1]}\cdot \chi _{\lbrack 2]}\cdot ...\cdot \chi
_{\lbrack p^{\prime }]},p^{\prime }\leq p,~\chi _{\lbrack i]}\in
T_{u}^{\ast }\xi _{N}.$ One defines a natural graded
d--algebra,%
\begin{equation}
grC_{u}=:\sum_{p}gr_{p}C_{u},~gr_{p}C_{u}=C_{u}^{(p)}/C_{u}^{(p-1)},
\label{pi3}
\end{equation}%
for which the natural projection is called the symbol map,%
\begin{equation*}
\sigma _{p}:C_{u}^{(p)}\rightarrow gr_{p}C_{u}.
\end{equation*}%
The natural graded d--algebra is canonical isomorphic to the
complexified exterior d--algebra $\Lambda _{\C}\left( T_{u}^{\ast
}\xi _{N}\right) ,$ the
isomorphism being defined as%
\begin{equation}
\Lambda _{\C}\left( T_{u}^{\ast }\xi _{N}\right) \ni \chi
_{\lbrack 1]}\wedge \chi _{\lbrack 2]}\wedge ...\wedge \chi
_{\lbrack p]}\rightarrow \sigma _{p}\left( \chi _{\lbrack
1]}\cdot \chi _{\lbrack 2]}\cdot ...\cdot \chi _{\lbrack
p]}\right) \in gr_{p}C_{u}.  \label{pi4}
\end{equation}

As a consequence of formulas (\ref{pi3}) and (\ref{pi4}), for a
canonical d--triple $\left[
\mathcal{A}_{[d]},\mathcal{H},D_{[d]}\right] $ over the
vector bundle $\xi _{N},$ one follows the property: a pair of operators $%
Q_{1}$ and $Q_{2}$ on $\mathcal{H}$ is of the form $Q_{1}=\pi
(\varpi )$ and
$Q_{2}=\pi (\delta \varpi )$ for some universal form $\varpi \in \Omega ^{p}%
\mathcal{A}_{[d]},$ if and only if there are sections $\rho _{1}$ of $%
C^{(p)} $ and $\rho _{2}$ of $C^{(p+1)}$ such that%
\begin{equation*}
Q_{1}=\gamma \left( \rho _{1}\right) \mbox{ and }Q_{2}=\gamma
\left( \rho _{2}\right)
\end{equation*}%
for which%
\begin{equation*}
\delta _{\xi _{N}}\sigma _{p}\left( \rho _{1}\right) =\sigma
_{p+1}\left( \rho _{2}\right) .
\end{equation*}

The introduced symbol map defines the canonical isomorphism
\begin{equation}
\sigma _{p}:\Omega _{D}^{p}\mathcal{A}_{[d]}\simeq \Gamma \left( \Lambda _{\C%
}^{p}T^{\ast }\xi _{N}\right)  \label{form1}
\end{equation}%
which commutes with the differential. With this isomorphism the
inner product on $\Omega _{D}^{p}\mathcal{A}_{[d]}$ (the scalar
product of forms) is proportional to the Riemannian inner product
of distinguished $p$--forms
on $\xi _{N},$%
\begin{equation}
<\varpi _{1},\varpi _{2}>_{p}=\left( -1\right) ^{p}\frac{2^{(n+m)/2+1-(n+m)}%
\pi ^{-(n+m)/2}}{(n+m)\Gamma \left( (n+m)/2\right) }\int_{\xi
_{N}}\varpi _{1}\wedge \ast \varpi _{2}  \label{form2}
\end{equation}%
for every $\varpi _{1},\varpi _{2}\in \Omega
_{D}^{p}\mathcal{A}_{[d]}\simeq \Gamma \left( \Lambda
_{\C}^{p}T^{\ast }\xi _{N}\right) .$

The proofs of formulas (\ref{form1}) and (\ref{form2}) are
similar to those given in \cite{landi} for $\xi _{N}=M.$

\subsection{Noncommutative Geometry and Anholonomic Gravity}

We introduce the concepts of generalized Lagrange and Finsler
geometry and outline the conditions when such structures can be
modeled on a Riemannian space by using anholnomic frames.

\subsubsection{Anisotropic spacetimes}

Different classes of commuative anisotropic spacetimes are
modeled by correponding parametriztions of some compatible (or
even non--compatible) N--connection, d--connection and d--metric
structrures on (pseudo) Riemannian spaces, tangent (or cotangent)
bundles, vector (or covector) bundles and their higher order
generalizations in their usual manifold,
supersymmetric, spinor, gauge like or another type approaches (see Refs. %
\cite{vexsol,miron,ma,bejancu,vspinors,vgauge,vmon1,vmon2}). Here
we revise the basic definitions and formulas which will be used
in further noncommutative embeddings and generalizations.

\paragraph{Anholonomic structures on Riemannian spaces:}

We can generate an anholonomic (equivalently, anisotropic)
structure on a Rieman space of dimension $(n+m)$ space $\ $(let
us denote this space $\ V^{(n+m)}$ and call it as a anholonomic
Riemannian space) by fixing an
anholonomic frame basis and co-basis with associated N--connection $%
N_{i}^{a}(x,y),$ respectively, as (\ref{dder}) and (\ref{ddif})
which splits the local coordinates $u^{\alpha }=(x^{i},y^{a})$
into two classes: $n$ holonomic coorinates, $x^{i},$ and $m$
anholonomic coordinates,$\,$\ $y^{a}.$ The d--metric
(\ref{dmetric}) on $V^{(n+m)}$,
\begin{equation}
G^{[R]}=g_{ij}(x,y)dx^{i}\otimes dx^{j}+g_{ab}(x,y)\delta
y^{a}\otimes \delta y^{b}  \label{dmetrr}
\end{equation}%
written with respect to a usual coordinate basis $du^{\alpha
}=\left(
dx^{i},dy^{a}\right) ,$%
\begin{equation*}
ds^{2}=\underline{g}_{\alpha \beta }\left( x,y\right) du^{\alpha
}du^{\beta }
\end{equation*}%
is a generic off--diagonal \ Riemannian metric parametrized as%
\begin{equation}
\underline{g}_{\alpha \beta }=\left[
\begin{array}{cc}
g_{ij}+N_{i}^{a}N_{j}^{b}g_{ab} & h_{ab}N_{i}^{a} \\
h_{ab}N_{j}^{b} & g_{ab}%
\end{array}%
\right] .  \label{odm}
\end{equation}%
Such type of metrics were largely investigated in the Kaluza--Klein gravity %
\cite{salam}, but also in the Einstein gravity \cite{vexsol}. An
off--diagonal metric (\ref{odm}) can be reduced to a block
$\left( n\times n\right) \oplus \left( m\times m\right) $ form
$\left( g_{ij},g_{ab}\right) , $ and even effectively
diagonalized in result of a superposition of ahnolonomic
N--transforms. It can be defined as an exact solution of the
Einstein equations. With respect to anholonomic frames, in
general, the Levi--Civita connection obtains a torsion component
(\ref{lcsym}). Every class of off--diagonal metrics can be
anholonomically equivalent to another ones for which it is not
possible to a select the Levi--Civita metric defied as the unique
torsionless and metric compatible linear connection. \ The
conclusion is that if anholonomic frames of reference, which
authomatically induce the torsion via anholonomy coefficients,
are considered on a Riemannian space we have to postulate
explicitly what type of linear connection (adapted both to the
anholonomic frame and metric structure) is chosen in order to
construct a Riemannian geometry and corresponding physical
models. For instance, we may postulate the connection
(\ref{lccon}) or the d--connection (\ref{dcon}). Both these
connections are metric compatible and transform into the usual
Christoffel symbols if the N--connection vanishes, i. e. the
local frames became holonomic. But, in general, anholonomic
frames and off--diagonal Riemannian metrics are connected with
anisotropic configurations which allow, in principle, to
model even Finsler like structures in (pseudo) Riemannian spaces \cite%
{vankin,vexsol}.

\paragraph{Finsler geometry and its almost Kahlerian model:}

The modern approaches to Finsler geometry are outlined in Refs. \cite%
{finsler,ma,miron,bejancu,vmon1,vmon2}. Here we emphasize that a
Finsler metric can be defined on a tangent bundle $TM$ with local coordinates $%
u^{\alpha }=(x^{i},y^{a}\rightarrow y^{i})$ of dimension $2n,$
with a d--metric (\ref{dmetric}) for which the Finsler metric, i.
e. the quadratic form
\begin{equation*}
g_{ij}^{[F]}=g_{ab}=\frac{1}{2}\frac{\partial ^{2}F^{2}}{\partial
y^{i}\partial y^{j}}
\end{equation*}%
is positive definite, is defined in this way: \ 1) A Finsler
metric on a real manifold $M$ is a function $F:TM\rightarrow \R$ which on $\widetilde{TM}%
=TM\backslash \{0\}$ is of class $C^{\infty }$ and $F$ is only
continuous on
the image of the null cross--sections in the tangent bundle to $M.$ 2) $%
F\left( x,\lambda y\right) =\lambda F\left( x,\lambda y\right) $ for every $%
\R_{+}^{\ast }.$ 3) The restriction of $F$ to $\widetilde{TM}$ is
a positive function. 4) $rank\left[ g_{ij}^{[F]}(x,y)\right] =n.$

The Finsler metric $F(x,y)$ and the quadratic form $g_{ij}^{[F]}$
can be used to define the Christoffel symbols (not those from the
usual Riemannian
geometry)%
\begin{equation*}
c_{jk}^{\iota }(x,y)=\frac{1}{2}g^{ih}\left( \partial
_{j}g_{hk}+\partial _{k}g_{jh}-\partial _{h}g_{jk}\right)
\end{equation*}%
which allows to define the Cartan nonlinear connection as
\begin{equation}
N_{j}^{i}(x,y)=\frac{1}{4}\frac{\partial }{\partial y^{j}}\left[
c_{lk}^{\iota }(x,y)y^{l}y^{k}\right]  \label{ncc}
\end{equation}%
where we may not distinguish the v- and h- indices taking on $TM$
the same values.

In Finsler geometry there were investigated different classes of
remarkable Finsler linear connections introduced by Cartan,
Berwald, Matsumoto and other ones (see details in Refs.
\cite{finsler,ma,bejancu}). Here we note
that we can introduce $g_{ij}^{[F]}=g_{ab}$ and $N_{j}^{i}(x,y)$ in (\ref%
{dmetric}) and construct a d--connection via formulas
(\ref{dcon}).

A usual Finsler space $F^{n}=\left( M,F\left( x,y\right) \right)
$ is completely defined by its fundamental tensor
$g_{ij}^{[F]}(x,y)$ and Cartan nonlinear connection
$N_{j}^{i}(x,y)$ and its chosen d--connection structure. But the
N--connection allows us to define an almost complex
structure $I$ on $TM$ as follows%
\begin{equation*}
I\left( \delta _{i}\right) =-\partial /\partial y^{i}\mbox{ and
}I\left(
\partial /\partial y^{i}\right) =\delta _{i}
\end{equation*}%
for which $I^{2}=-1.$

The pair $\left( G^{[F]},I\right) $ consisting from a Riemannian metric on $%
TM,$%
\begin{equation}
G^{[F]}=g_{ij}^{[F]}(x,y)dx^{i}\otimes
dx^{j}+g_{ij}^{[F]}(x,y)\delta y^{i}\otimes \delta y^{j}
\label{dmetricf}
\end{equation}%
and the almost complex structure $I$ defines an almost Hermitian
structure
on $\widetilde{TM}$ associated to a 2--form%
\begin{equation*}
\theta =g_{ij}^{[F]}(x,y)\delta y^{i}\wedge dx^{j}.
\end{equation*}%
This model of Finsler geometry is called almost Hermitian and denoted $%
H^{2n} $ and it is proven \cite{ma} that is almost Kahlerian, i.
e. the form
$\theta $ is closed. The almost Kahlerian space $K^{2n}=\left( \widetilde{TM}%
,G^{[F]},I\right) $ is also called the almost Kahlerian model of
the Finsler space $F^{n}.$

On Finsler (and their almost Kahlerian models) spaces one
distinguishes the almost Kahler linear connection of Finsler
type, $D^{[I]}$ on $\widetilde{TM} $ with the property that this
covariant derivation preserves by parallelism the vertical
distribution and is compatible with the almost Kahler structure
$\left( G^{[F]},I\right) ,$ i.e.
\begin{equation*}
D_{X}^{[I]}G^{[F]}=0\mbox{ and }D_{X}^{[I]}I=0
\end{equation*}%
for \ every d--vector field on $\widetilde{TM}.$ This
d--connection is defined by the data
\begin{equation*}
\Gamma =\left(
L_{jk}^{i},L_{bk}^{a}=0,C_{ja}^{i}=0,C_{bc}^{a}\rightarrow
C_{jk}^{i}\right)
\end{equation*}%
with $L_{jk}^{i}$ and $C_{jk}^{i}$ computed as in the formulas
(\ref{dcon}) by using $g_{ij}^{[F]}$ and $N_{j}^{i}$ from
(\ref{ncc}).

We emphasize that a Finsler space $F^{n}$ with a d--metric
(\ref{dmetricf}) and Cartan's N--connection structure
(\ref{ncc}), or the corresponding almost Hermitian (Kahler) model
$H^{2n},$ can be equivalently modeled on a Riemannian space of
dimension $2n$ provided with an off--diagonal Riemannian metric
(\ref{odm}). From this viewpoint a Finsler geometry is a
corresponding Riemannian geometry with a respective off--diagonal
metric (or, equivalently, with an anholonomic frame structure
with associated N--connection) and a corresponding prescription
for the type of linear connection chosen to be compatible with
the metric and N--connection structures.

\paragraph{Lagrange and generalized Lagrange geometry:}

The Lagrange spaces were introduced in order to generalize the
fundamental
concepts in mechanics \cite{kern} and investigated in Refs. \cite{ma} (see %
\cite{vspinors,vgauge,vsuper,vstr2,vmon1,vmon2} for their spinor,
gauge and supersymmetric generalizations).

A Lagrange space $L^{n}=\left( M,L\left( x,y\right) \right) $ is
defined as a pair which consists of a real, smooth
$n$--dimensional manifold $M$ and regular Lagrangian
$L:TM\rightarrow \R.$ Similarly as for Finsler spaces one
introduces the symmetric d--tensor field
\begin{equation}
g_{ij}^{[L]}=g_{ab}=\frac{1}{2}\frac{\partial ^{2}L}{\partial
y^{i}\partial y^{j}}.  \label{mfl}
\end{equation}%
So, the Lagrangian $L(x,y)$ is like the square of the fundamental
Finsler metric, $F^{2}(x,y),$ but not subjected to any
homogeneity conditions.

In the rest me can introduce similar concepts of almost Hermitian
(Kahlerian) models of Lagrange spaces as for the Finsler spaces,
by using the similar definitions and formulas as in the previous
subsection, but changing $g_{ij}^{[F]}\rightarrow g_{ij}^{[L]}.$

R. Miron introduced the concept of generalized Lagrange space,
GL--space (see details in \cite{ma}) and a corresponding
N--connection geometry on $TM$ when the fundamental metric
function $g_{ij}=g_{ij}\left( x,y\right) $ is a general one, not
obligatory defined as a second derivative from a Lagrangian as in
(\ref{mfl}). The corresponding almost Hermitian (Kahlerian)
models of GL--spaces were investigated and applied in order to
elaborate generalizations of gravity and gauge theories
\cite{ma,vgauge}.

Finally, a few remarks on definition of gravity models with
generic local anisotropy on anholonomic Riemannian, Finsler or
(generalized) Lagrange spaces and vector bundles. So, by choosing
a d-metric (\ref{dmetric}) (in particular cases (\ref{dmetrr}),
or (\ref{dmetricf}) with $g_{ij}^{[F]},$ or $g_{ij}^{[L]})$ we
may compute the coefficients of, for instance, d--connection
(\ref{dcon}), d--torsion (\ref{dtors}) and (\ref{dcurvatures})
and even to write down the explicit form of Einstein equations (\ref%
{einsteq2}) which define such geometries. For instance, in a series of works %
\cite{vankin,vexsol,vmon2} we found explicit solutions when
Finsler like and another type anisotropic configurations are
modeled in anisotropic kinetic theory and irreversible
thermodynamics and even in Einstein or
low/extra--dimension gravity as exact solutions of the vacuum (\ref{einsteq2}%
) and nonvacuum (\ref{einsteq3}) Einstein equations. From the
viewpoint of the geometry of anholonomic frames is not much
difference between the usual Riemannian geometry and its Finsler
like generalizations. The explicit form and parametrizations of
coefficients of metric, linear connections, torsions, curvatures
and Einstein equations in all types of mentioned geometric models
depends on the type of anholomic frame relations and
compatibility metric conditions between the associated
N--connection structure and linear connections we fixed. Such
structures can be correspondingly picked up from a noncommutative
functional model, for instance from some almost Hermitian
structures over projective modules and/or generalized to some
noncommutative configurations.

\subsection{Noncommutative Finsler like gravity models}

We shall briefly describe two possible approaches to the
construction of gravity models with generic anisotropy following
from noncommutative geometry which while agreeing for the
canonical d--triples associated with vector bundles provided with
N--connection structure. Because in the previous section we
proved that the Finsler geometry and its extensions are
effectively modeled by anholonomic structures on Riemannian
manifolds (bundles) we shall only emphasize the basic ideas how
from the beautiful result by Connes \cite{connes,connes1} we may
select an anisotropic gravity (possible alternative approaches to
noncommutative gravity are examined in Refs.
\cite{ch1,madore,hawkins,landi1,landi2}; by introducing
anholonomic frames with associated N--connections those models
also can be transformed into certain anisotropic ones; we omit
such considerations in the present work).

\subsubsection{Anisotropic gravity a la Connes--Deximier--Wodzicki}

The first scheme to construct gravity models in noncommutative
geometry (see details in \cite{connes,landi}) may be extend for
vector bundles provided with N--connection structure (i. e. to
projective finite distinguished moduli) and in fact to
reconstruct the full anisotorpic (for instance, Finsler) geometry
from corresponding distinguishing of the Diximier trace and the
Wodzicki residue.

Let us consider a smooth compact vector bundle $\xi _{N}$ without
boundary and of dimension $n+m$ and $D_{[t]}$ as a ''symbol'' for
a time being operator and denote $\mathcal{A}_{[d]}=C^{\infty
}\left( \xi _{N}\right) .$ For a unitary representation $\left[
\mathcal{A}_{\pi },D_{\pi }\right] $ $\ $of the couple $\left(
\mathcal{A}_{[d]},D_{[t]}\right) $ as operators on an
Hilbert space $\mathcal{H}_{\pi }$ provided with a real structure operator $%
J_{\pi },$ such that $\left[ \mathcal{A}_{\pi },D_{\pi },\mathcal{H},J_{\pi }%
\right] $ $\ $satisfy all axioms of a real spectral d--triple.
Then, one holds the properties:

\begin{enumerate}
\item There is a unique Riemannian d--metric $g_{\pi }$ on $\xi _{N}$ such
the geodesic distance in the total space of the vector bundle
between every
two points $u_{[1]}$ and $u_{[2]}$ is given by%
\begin{equation*}
d\left( u_{[1]},u_{[2]}\right) =\sup_{a\in
\mathcal{A}_{[d]}}\left\{ \left| a(u_{[1]})-a(u_{[2]})\right|
:\left\| D_{\pi },\pi \left( a\right) \right\|
_{\mathcal{B}(\mathcal{H}_{\pi })}\leq 1\right\} .
\end{equation*}

\item The d--metric $g_{\pi }$ depends only on the unitary equivalence class
of the representations $\pi .$ The fibers of the map $\pi
\rightarrow g_{\pi }$ form unitary equivalence classes of
representations to metrics define a finite collection of affine
spaces $\mathcal{A}_{\sigma }$ parametrized by the spin
structures $\sigma $ on $\xi _{N}.$ These spin structures depends
on the type of d--metrics we are using in $\xi _{N}.$

\item The action functional given by the Diximier trace%
\begin{equation*}
G\left( D_{[t]}\right) =tr_{\varpi }\left( D_{[t]}^{n+m-2}\right)
\end{equation*}%
is a positive quadratic d--form with a unique minimum $\pi
_{\sigma }$ for each $\mathcal{A}_{\sigma }.$ At the minimum, the
values of $G\left( D_{[t]}\right) $ coincides with the Wodzicki
residue of $D_{\sigma }^{n+m-2}$ and is proportional to the
Hilbert--Einstein action for a fixed
d--connection,%
\begin{eqnarray*}
G\left( D_{\sigma }\right) ={Re}s_{W}\left( D_{\sigma
}^{n+m-2}\right)
&=&:\frac{1}{\left( n+m\right) \left( 2\pi \right) ^{n+m}}%
\int\limits_{S^{\ast }\xi _{N}}tr\left[ \sigma _{-(n+m)}\left(
u,u^{\prime
}\right) \delta u\delta u^{\prime }\right] \\
&=&c_{n+m}\int\limits_{\xi _{N}}\overleftarrow{R}\delta u,
\end{eqnarray*}%
where%
\begin{equation*}
c_{n+m}=\frac{n+m-2}{12}\frac{2^{[(n+m)/2]}}{\left( 4\pi \right)
^{(n+m)/2}\Gamma \left( \frac{n+m}{2}+1\right) },
\end{equation*}%
$\sigma _{-(n+m)}\left( u,u^{\prime }\right) $ is the part of
order $-(n+m)$ of the total symbol of $D_{\sigma }^{n+m-2},$
$\overleftarrow{R}$ is the scalar curvature (\ref{dscalar}) on
$\xi _{N}$ and $tr$ is a normalized Clifford trace.

\item It is defined a representation of $\left( \mathcal{A}%
_{[d]},D_{[t]}\right) $ for every minimum $\pi _{\sigma }$ on the
Hilbert space of square integrable d--spinors
$\mathcal{H}=L^{2}(\xi _{N},S_{\sigma
})$ where $\mathcal{A}_{[d]}$ acts by multiplicative operators and $%
D_{\sigma }$ is the Dirac operator of chosen d--connection. If
there is no real structure $J,$ one has to replace $spin\,\ $by
$spin^{c}$ (for d--spinors investigated in Refs.
\cite{vspinors,vmon1,vmon2}). In this case there is not \ a
uniqueness and the minimum of the functional $G\left(
D\right) $ is reached on a linear subspace of $\mathcal{A}_{\sigma }$ with $%
\sigma $ a fixed $spin^{c}$ structure. This subspace is parametrized by the $%
U\left( 1\right) $ gauge potentials entering in the $spin^{c}$
Dirac operator (the rest properties hold).
\end{enumerate}

The properties 1-4 are proved in a similar form as in \cite%
{kalau,hawkins,landi}, but all computations are distinguished by
the N--connection structure and a fixed type of d--connection (we
omit such details). We can generate an anholonomic Riemannian,
Finsler or Lagrange
gravity depending on the class of d--metrics ((\ref{dmetrr}), (\ref{dmetricf}%
), (\ref{mfl}), or a general one for vector bundles
(\ref{dmetric})) we choose.

\subsubsection{Spectral anisotropic Gravity}

Consider a canonical d--triple $\left[
\mathcal{A}_{[d]}=C^{\infty }\left( \xi _{N}\right)
,\mathcal{H}=L^{2}\left( \xi _{N}\right) ,_{[d]}D\right] $ \
defined in subsection \ref{spinorsubsection} for a vector bundle
$\xi _{N},$ where $_{[d]}D$ is the Dirac d--operator
(\ref{dirac}) defined for a d--connection on \ $\xi _{N}.$ We are
going to compute the action
\begin{equation}
S_{G}\left( _{[d]}D,\Lambda \right) =tr_{\mathcal{H}}\left[ \chi
\left( \frac{_{\lbrack d]}D^{2}}{\Lambda ^{2}}\right) \right] ,
\label{action}
\end{equation}%
depending on the spectrum of $_{[d]}D,$ were $tr_{\mathcal{H}}$
is the usual trace in the Hilbert space, $\Lambda $ is the cutoff
parameter and $\chi $ will be closed as a suitable cutoff
function which cut off all eigenvalues of $_{[d]}D^{2}$ larger
than $\Lambda ^{2}.$ By using the Lichnerowicz formula, in our
case with operators for a vector bundle, and the keat kernel
expansion (similarly as for the proof summarized in Ref.
\cite{landi})
\begin{equation*}
S_{G}\left( _{[d]}D,\Lambda \right) =\sum_{k\geq
0}f_{k}a_{k}\left( _{[d]}D^{2}/\Lambda ^{2}\right) ,
\end{equation*}%
were the coefficients $f_{k}$ are computed
\begin{equation*}
f_{0}=\int\limits_{0}^{\infty }\chi \left( z\right)
zdz,~~f_{2}=\int\limits_{0}^{\infty }\chi \left( z\right)
dz,~~~f_{2(k^{\prime }+2)}=\left( -1\right) ^{k^{\prime }}\chi
^{(k^{\prime })}\left( 0\right) ,k^{\prime }\geq 0,
\end{equation*}%
$\chi ^{(k^{\prime })}$ denotes the $k^{\prime }$th derivative on
its
argument, the so--called non--vanishing Seeley--de Witt coefficients $%
a_{k}\left( _{[d]}D^{2}/\Lambda ^{2}\right) $ are defined for
even values of
$k$ as integrals%
\begin{equation*}
a_{k}\left( _{[d]}D^{2}/\Lambda ^{2}\right) =\int\limits_{\xi
_{N}}a_{k}\left( u;_{[d]}D^{2}/\Lambda ^{2}\right) \sqrt{g}\delta
u
\end{equation*}%
with the first three subintegral functions given by%
\begin{eqnarray*}
a_{0}\left( u;_{[d]}D^{2}/\Lambda ^{2}\right) &=&\Lambda
^{4}\left( 4\pi
\right) ^{-(n+m)/2}trI_{2^{[(n+m)/2]}}, \\
a_{2}\left( u;_{[d]}D^{2}/\Lambda ^{2}\right) &=&\Lambda
^{2}\left( 4\pi \right) ^{-(n+m)/2}\left(
-\overleftarrow{R}/6+E\right) trI_{2^{[(n+m)/2]}},
\\
a_{4}\left( u;_{[d]}D^{2}/\Lambda ^{2}\right) &=&\left( 4\pi
\right)
^{-(n+m)/2}\frac{1}{360}(-12D_{\mu }D^{\mu }\overleftarrow{R}+5%
\overleftarrow{R}^{2}-2R_{\mu \nu }R^{\mu \nu } \\
&&-\frac{7}{4}R_{\mu \nu \alpha \beta }R^{\mu \nu \alpha \beta
}-60RE+180E^{2}+60D_{\mu }D^{\mu
}\overleftarrow{E})trI_{2^{[(n+m)/2]}},
\end{eqnarray*}%
and $\overleftarrow{E}=:_{[d]}D^{2}-\bigtriangledown ^{\lbrack S]}={%
\overleftarrow{R}/4},$ see (\ref{lich}). We can use for the
function $\chi $ the characteristic value of the interval
$[0,1],$ namely $\chi \left( z\right) =1$ for $z\leq 1$ and $\chi
\left( z\right) =0$ for $z\geq 1,$
possibly 'smoothed out' at $z=1,$ we get%
\begin{equation*}
f_{0}=1/2,f_{2}=1,f_{2(k^{\prime }+2)}=0,k^{\prime }\geq 0.
\end{equation*}

We compute (a similar calculus is given in \cite{landi}; we only
distinguish
the curvature scalar, the Ricci and curvature d--tensor) the action (\ref%
{action}),%
\begin{equation*}
S_{G}\left( _{[d]}D,\Lambda \right) =\Lambda
^{4}\frac{2^{(n+m)/2-1}}{\left( 4\pi \right)
^{(n+m)/2}}\int\limits_{\xi _{N}}\sqrt{g}\delta u+\frac{\Lambda
^{2}}{6}\frac{2^{(n+m)/2-1}}{\left( 4\pi \right)
^{(n+m)/2}}\int\limits_{\xi _{N}}\sqrt{g}\overleftarrow{R}\delta
u.
\end{equation*}%
This action is dominated by the first term with a huge
cosmological
constant. But this constant can be eliminated \cite{landi2} if the function $%
\chi \left( z\right) $ is replaced by $\widetilde{\chi }\left(
z\right) =\chi \left( z\right) -\alpha \chi \left( \beta z\right)
$ with any two numbers $\alpha $ and $\beta $ such that $\alpha
=\beta ^{2}$ and $\beta
\geq 0,\beta \neq 1.$ The final form of the action becomes%
\begin{equation}
S_{G}\left( _{[d]}D,\Lambda \right) =\left( 1-\frac{\alpha }{\beta ^{2}}%
\right) f_{2}\frac{\Lambda ^{2}}{6}\frac{2^{(n+m)/2-1}}{\left(
4\pi \right) ^{(n+m)/2}}\int\limits_{\xi
_{N}}\sqrt{g}\overleftarrow{R}\delta u+O\left( (\Lambda
^{2})^{0}\right) .  \label{act1}
\end{equation}

From the action (\ref{act1}) we can generate different models of
anholonomic Riemannian, Finsler or Lagrange gravity depending on
the class of d--metrics ((\ref{dmetrr}), (\ref{dmetricf}),
(\ref{mfl}), or a general one for vector bundles (\ref{dmetric}))
we parametrize for computations. But this construction has a
problem connected with ''spectral invariance versus
diffeomorphysms invariance on manifolds or vector bundles. Let us denote by $%
spec\left( \xi _{N},_{[d]}D\right) $ the spectrum of the Dirac
d--operator with each eigenvalue repeated according to its
multiplicity. Two vector
bundles $\xi _{N}$ and $\xi _{N}^{\prime }$ are called isospectral if $%
spec\left( \xi _{N},_{[d]}D\right) =spec\left( \xi _{N}^{\prime
},_{[d]}D\right) ,$ which defines an invariant transform of the action (\ref%
{action}). There are manifolds (and in consequence vector
bundles) which are isospectral without being isometric (the
converse is obviously true). This is known as a fact that one
cannot 'hear the shape of a drum \cite{drum} because the spectral
invariance is stronger that usual difeomorphism invariance.

In spirit of spectral gravity, the eigenvalues of the Dirac
operator are diffeomorphic invariant functions of the geometry
and therefore true observable in general relativity. As we have
shown in this section they can be taken as a set of variables for
invariant descriptions to the anholonomic dynamics of the
gravitational field with (or not) local anisotropy in different
approaches of anholonomic Riemannian gravity and Finsler like
generalizations. But in another turn there exist isospectral
vector bundles which fail to be isometric. Thus, the eigenvalues
of the Dirac operator cannot be used to distinguish among such
vector bundles (or manifolds). A rigorous analysis is also
connected with the type of d--metric and d--connection structures
we prescribe for our geometric and physical models.

Finally, we remark that there are different models of gravity with
noncommutative setting (see, for instance, Refs. \cite%
{ch1,madore,hawkins,landi1,landi2,cl,kalau}). By introducing
nonlinear connections in a respective commutative or
noncommutative variant we can transform such theories to be
anholonomic, i. e. locally anisotropic, in different approaches
with (pseudo) Riemannian geometry and Finsler/Lagrange or
Hamilton extensions.

\section{Noncommutative Finsler--Gauge Theories}

The bulk of noncommutative models extending both locally
isotropic and anisotropic gravity theories are confrunted with
the problem of definition of noncommutative variants of
pseudo--Eucliedean and pseudo--Riemannian metrics. The problem is
connected with the fact of generation of noncommutative metric
structures via the Moyal results in complex and noncommutative
metrics. In order to avoid this difficulty we elaborated a model
of noncommutative gauge gravity (containing as particular case the
Einstein general relativity theory) starting from a variant of
gauge gravity being equivalent to the Einstein gravity and
emphasizing in a such approach the tetradic (frame) and
connection structures, but not the metric configuration (see
Refs. \cite{vnonc}). The metric for such theories is induced from
the frame structure which can be holonomic or anholonomic. The
aim of this section is to generalize our results on
noncommutative gauge gravity as to include also possible
anisotropies in different variants of gauge realization of
anholonomic Einstein and Finsler like generalizations formally
developed in Refs. \cite{vgauge,vmon1,vmon2}.

A still presented drawback of noncommutative geometry and physics
is that there is not yet formulated a generally accepted approach
to interactions of elementary particles coupled to gravity. There
are improved Connes--Lott and Chamsedine--Connes models of
nocommutative geometry \cite{connes1,cl} which yielded action
functionals typing together the gravitational and Yang--Mills
interactions and gauge bosons the Higgs sector (see also the approaches \cite%
{hawkins} and, for an outline of recent results, \cite{majid}).

In the last years much work has been made in noncommutative
extensions of
physical theories (see reviews and original results in Refs. \cite%
{douglas,soch}). It was not possible to formulate gauge theories
on noncommutative spaces \cite{cds,sw,js,mssw} with Lie algebra
valued infinitesimal transformations and with Lie algebra valued
gauge fields. In order to avoid the problem it was suggested to
use enveloping algebras of the Lie algebras for setting this type
of gauge theories and showed that in spite of the fact that such
enveloping algebras are infinite--dimensional one can restrict
them in a way that it would be a dependence on the Lie algebra
valued parameters, the Lie algebra valued gauge fields and their
spacetime derivatives only.

We follow the method of restricted enveloping algebras \cite{js}
and construct gauge gravitational theories by stating
corresponding structures with semisimple or nonsemisimple Lie
algebras and their extensions. We consider power series of
generators for the affine and non linear realized de Sitter gauge
groups and compute the coefficient functions of all the higher
powers of the generators of the gauge group which are functions of
the coefficients of the first power. Such constructions are based
on the Seiberg--Witten map \cite{sw} and on the formalism of
$\ast $--product formulation of the algebra \cite{w} when for
functional objects, being functions of commuting variables, there
are associated some algebraic noncommutative properties encoded
in the $\ast $--product.

The concept of gauge theory on noncommutative spaces was
introduced in a geometric manner \cite{mssw} by defining the
covariant coordinates without speaking about derivatives and this
formalism was developed for quantum planes \cite{wz}. In this
section we shall prove the existence for noncommutative spaces of
gauge models of gravity which agrees with usual gauge gravity
theories being equivalent, or extending, the general relativity
theory (see works \cite{pd,ts} for locally isotropic and
anisotropic spaces and corresponding reformulations and
generalizations respectively for anholonomic frames \cite{vd} and
locally anisotropic (super) spaces
\cite{vgauge,vsuper,vstring,vmon1}) in the limit of commuting
spaces.

\subsection{Star--products and enveloping algebras in noncommutative spaces}

For a noncommutative space the coordinates ${\hat u}^i,$
$(i=1,...,N)$ satisfy some noncommutative relations
\begin{equation}  \label{ncr}
[{\hat u}^i,{\hat u}^j]=\left\{
\begin{array}{rcl}
& i\theta ^{ij}, & \theta ^{ij}\in \C,\mbox{ canonical structure;
} \\
& if_k^{ij}{\hat u}^k, & f_k^{ij}\in \C,\mbox{ Lie
structure; } \\
& iC_{kl}^{ij}{\hat u}^k{\hat u}^l, & C_{kl}^{ij}\in \C ,%
\mbox{ quantum
plane }%
\end{array}
\right.
\end{equation}
where $\C$ denotes the complex number field.

The noncommutative space is modeled as the associative algebra of
$\C;$\ this algebra is freely generated by the coordinates modulo
ideal $\mathcal{R}
$ generated by the relations (one accepts formal power series)\ $\mathcal{A}%
_{u}=\C[[{\hat u}^1,...,{\hat u}^N]]/\mathcal{R}.$ One restricts attention %
\cite{jssw} to algebras having the (so--called,
Poincare--Birkhoff--Witt) property that any element of
$\mathcal{A}_{u}$ is defined by its coefficient
function and vice versa,%
\begin{equation*}
\widehat{f}=\sum\limits_{L=0}^{\infty }f_{i_{1},...,i_{L}}:{\hat{u}}%
^{i_{1}}\ldots {\hat{u}}^{i_{L}}:\quad \mbox{ when
}\widehat{f}\sim \left\{ f_{i}\right\} ,
\end{equation*}%
where $:{\hat{u}}^{i_{1}}\ldots {\hat{u}}^{i_{L}}:$ denotes that
the basis
elements satisfy some prescribed order (for instance, the normal order $%
i_{1}\leq i_{2}\leq \ldots \leq i_{L},$ or, another example, are
totally symmetric). The algebraic properties are all encoded in
the so--called diamond $(\diamond )$ product which is defined by
\begin{equation*}
\widehat{f}\widehat{g}=\widehat{h}~\sim ~\left\{ f_{i}\right\}
\diamond \left\{ g_{i}\right\} =\left\{ h_{i}\right\} .
\end{equation*}

In the mentioned approach to every function $f(u)=f(u^{1},\ldots
,u^{N})$ of commuting variables $u^{1},\ldots ,u^{N}$ one
associates an element of algebra $\widehat{f}$ when the commuting
variables are substituted by anticommuting ones,
\begin{equation}
f(u)=\sum f_{i_{1}\ldots i_{L}}u^{1}\cdots u^{N}\rightarrow \widehat{f}%
=\sum\limits_{L=0}^{\infty }f_{i_{1},...,i_{L}}:{\hat{u}}^{i_{1}}\ldots {%
\hat{u}}^{i_{L}}:  \notag
\end{equation}%
when the $\diamond $--product leads to a bilinear $\ast
$--product of
functions (see details in \cite{mssw})%
\begin{equation*}
\left\{ f_{i}\right\} \diamond \left\{ g_{i}\right\} =\left\{
h_{i}\right\} \sim \left( f\ast g\right) \left( u\right) =h\left(
u\right) .
\end{equation*}

The $*$--product is defined respectively for the cases (\ref{ncr})
\begin{equation*}
f*g=\left\{
\begin{array}{rcl}
\exp [{\frac i2}{\frac \partial {\partial u^i}}{\theta
}^{ij}\frac \partial {\partial {u^{\prime }}^j}]f(u)g(u^{\prime
}){|}_{u^{\prime }\to u}, &  &
\\
\exp [\frac i2u^kg_k(i\frac \partial {\partial u^{\prime
}},i\frac \partial
{\partial u^{\prime \prime }})]f(u^{\prime })g(u^{\prime \prime }){|}%
_{u^{\prime \prime }\to u}^{u^{\prime }\to u}, &  &  \\
q^{{\frac 12}(-u^{\prime }{\frac \partial {\partial u^{\prime
}}}v{\frac
\partial {\partial v}}+u{\frac \partial {\partial u}}v^{\prime }{\frac
\partial {\partial v^{\prime }}})}f(u,v)g(u^{\prime },v^{\prime }){|}%
_{v^{\prime }\to v}^{u^{\prime }\to u}, &  &
\end{array}
\right.
\end{equation*}
where there are considered values of type%
\begin{eqnarray}
&e^{ik_n\widehat{u}^n}& e^{ip_{nl}\widehat{u}^n}
=e^{i\{k_n+p_n+\frac
12g_n\left( k,p\right) \}\widehat{u}^n,}  \label{gdecomp} \\
&g_n\left( k,p\right)& = -k_ip_jf_{\ n}^{ij}+\frac 16k_ip_j\left(
p_k-k_k\right) f_{\ m}^{ij}f_{\ n}^{mk}+...,  \notag \\
&e^Ae^B& = e^{A+B+\frac 12[A,B]+\frac 1{12}\left(
[A,[A,B]]+[B,[B,A]]\right) }+...  \notag
\end{eqnarray}
and for the coordinates on quantum (Manin) planes one holds the relation $%
uv=qvu.$

A non--abelian gauge theory on a noncommutative space is given by
two algebraic structures, the algebra $\mathcal{A}_{u}$ and a
non--abelian Lie
algebra $\mathcal{A}_{I}$ of the gauge group with generators $%
I^{1},...,I^{S} $ and the relations
\begin{equation}
\lbrack I^{\underline{s}},I^{\underline{p}}]=if_{~\underline{t}}^{\underline{%
s}\underline{p}}I^{\underline{t}}.  \label{commutators1}
\end{equation}%
In this case both algebras are treated on the same footing and
one denotes
the generating elements of the big algebra by $\widehat{u}^{i},$%
\begin{equation}
\widehat{z}^{\underline{i}}=\{\widehat{u}^{1},...,\widehat{u}%
^{N},I^{1},...,I^{S}\},\mathcal{A}_{z}=\C[[\widehat{u}^1,...,%
\widehat{u}^{N+S}]]/\mathcal{R}  \notag
\end{equation}%
and the $\ast $--product formalism is to be applied for the whole algebra $%
\mathcal{A}_{z}$ when there are considered functions of the
commuting variables $u^{i}\ (i,j,k,...=1,...,N)$ and
$I^{\underline{s}}\ (s,p,...=1,...,S).$

For instance, in the case of a canonical structure for the space variables $%
u^{i}$ we have
\begin{eqnarray}
(F\ast G)(u) &=&e^{\frac{i}{2}\left( \theta ^{ij}\frac{\partial
}{\partial u^{\prime i}}\frac{\partial }{\partial u^{\prime
\prime j}}+t^{s}g_{s}\left( i\frac{\partial }{\partial t^{\prime
}},i\frac{\partial }{\partial t^{\prime \prime }}\right) \right)
\times F\left( u^{\prime },t^{\prime }\right) G\left( u^{\prime
\prime },t^{\prime \prime }\right) \mid _{t^{\prime }\rightarrow
t,t^{\prime \prime }\rightarrow t}^{u^{\prime }\rightarrow
u,u^{\prime \prime }\rightarrow u}.}  \label{csp1} \\
&&  \notag
\end{eqnarray}%
This formalism was developed in \cite{jssw} for general Lie
algebras. In this section we consider those cases when in the
commuting limit one obtains the gauge gravity and general
relativity theories or some theirs anisotropic generalizations..

\subsection{Enveloping algebras for gauge gravity connections}

In order to construct gauge gravity theories on noncommutative
space we define the gauge fields as elements the algebra
$\mathcal{A}_{u}$ that form representation of the generator
$I$--algebra for the de Sitter gauge group. For commutative
spaces it is known \cite{pd,ts,vgauge} that an equivalent
re--expression of the Einstein theory as a gauge like theory
implies, for both locally isotropic and anisotropic spacetimes,
the nonsemisimplicity of the gauge group, which leads to a
nonvariational theory in the total space of the bundle of locally
adapted affine frames (to this class one belong the gauge
Poincare theories;\ on metric--affine and gauge gravity models see
original results and reviews in \cite{ut}). By using auxiliary
biliniear forms, instead of degenerated Killing form for the
affine structural group, on fiber spaces, the gauge models of
gravity can be formulated to be variational. After projection on
the base spacetime, for the so--called Cartan connection form,
the Yang--Mills equations transforms equivalently into the
Einstein equations for general relativity \cite{pd}. A variational
gauge gravitational theory can be also formulated by using a
minimal
extension of the affine structural group ${\mathcal{A}f}_{3+1}\left( {\R}%
\right) $ to the de Sitter gauge group $S_{10}=SO\left(
4+1\right) $ acting on ${\R}^{4+1}$ space.

\subsubsection{Nonlinear gauge theories of de Sitter group in commutative
spaces}

Let us consider the de Sitter space $\Sigma ^{4}$ as a
hypersurface given by the equations $\eta _{AB}u^{A}u^{B}=-l^{2}$
in the four dimensional flat space enabled with diagonal metric
$\eta _{AB},\eta _{AA}=\pm 1$ (in this section
$A,B,C,...=1,2,...,5),$ where $\{u^{A}\}$ are global Cartesian
coordinates in $\R^{5};l>0$ is the curvature of de Sitter space.
The de Sitter group $S_{\left( \eta \right) }=SO_{\left( \eta
\right) }\left( 5\right) $ is defined as the isometry group of
$\Sigma ^{5}$--space with $6$ generators of Lie algebra
${\mathit{s}o}_{\left( \eta \right) }\left( 5\right) $ satisfying
the commutation relations
\begin{eqnarray}
\left[ M_{AB},M_{CD}\right] &=&\eta _{AC}M_{BD}-\eta
_{BC}M_{AD}-\eta
_{AD}M_{BC}+\eta _{BD}M_{AC}.  \label{dsc} \\
&&  \notag
\end{eqnarray}

Decomposing indices $A,B,...$ as $A=\left( \underline{\alpha
},5\right)
,B=\left( \underline{\beta },5\right) ,...,$ the metric $\eta _{AB}$ as $%
\eta _{AB}=\left( \eta _{\underline{\alpha }\underline{\beta
}},\eta
_{55}\right) ,$ and operators $M_{AB}$ as $M_{\underline{\alpha }\underline{%
\beta }}=\mathcal{F}_{\underline{\alpha }\underline{\beta }}$ and $P_{%
\underline{\alpha }}=l^{-1}M_{5\underline{\alpha }},$ we can write (\ref{dsc}%
) as
\begin{eqnarray}
\left[ \mathcal{F}_{\underline{\alpha }\underline{\beta }},\mathcal{F}_{%
\underline{\gamma }\underline{\delta }}\right] &=&\eta _{\underline{\alpha }%
\underline{\gamma }}\mathcal{F}_{\underline{\beta
}\underline{\delta }}-\eta
_{\underline{\beta }\underline{\gamma }}\mathcal{F}_{\underline{\alpha }%
\underline{\delta }}+\eta _{\underline{\beta }\underline{\delta }}\mathcal{F}%
_{\underline{\alpha }\underline{\gamma }}-\eta _{\underline{\alpha }%
\underline{\delta }}\mathcal{F}_{\underline{\beta
}\underline{\gamma }},
\notag \\
\left[ P_{\underline{\alpha }},P_{\underline{\beta }}\right] &=&-l^{-2}%
\mathcal{F}_{\underline{\alpha }\underline{\beta }},\left[ P_{\underline{%
\alpha }},\mathcal{F}_{\underline{\beta }\underline{\gamma }}\right] =\eta _{%
\underline{\alpha }\underline{\beta }}P_{\underline{\gamma }}-\eta _{%
\underline{\alpha }\underline{\gamma }}P_{\underline{\beta }},
\label{dsca}
\\
&&  \notag
\end{eqnarray}%
where we decompose the Lie algebra ${\mathit{s}o}_{\left( \eta
\right) }\left( 5\right) $ into a direct sum,
${\mathit{s}o}_{\left( \eta \right) }\left( 5\right)
={\mathit{s}o}_{\left( \eta \right) }(4)\oplus V_{4},$ where
$V_{4}$ is the vector space stretched on vectors
$P_{\underline{\alpha }}.$ We remark that $\Sigma ^{4}=S_{\left(
\eta \right) }/L_{\left( \eta \right) },$ where $L_{\left( \eta
\right) }=SO_{\left( \eta \right) }\left(
4\right) .$ For $\eta _{AB}=diag\left( 1,-1,-1,-1\right) $ and $%
S_{10}=SO\left( 1,4\right) ,L_{6}=SO\left( 1,3\right) $ is the
group of Lorentz rotations.

In this paper the generators $I^{\underline{a}}$ and structure constants $%
f_{~\underline{t}}^{\underline{s}\underline{p}}$ from
(\ref{commutators1})
are parametrized just to obtain de Sitter generators and commutations (\ref%
{dsca}).

The action of the group $S_{\left( \eta \right) }$ can be realized by using $%
4\times 4$ matrices with a parametrization distinguishing the subgroup $%
L_{\left( \eta \right) }:$%
\begin{equation}  \label{parametriz}
B=bB_L,
\end{equation}
where%
\begin{equation*}
B_L=\left(
\begin{array}{cc}
L & 0 \\
0 & 1%
\end{array}
\right) ,
\end{equation*}
$L\in L_{\left( \eta \right) }$ is the de Sitter bust matrix
transforming the vector $\left( 0,0,...,\rho \right) \in {\R}^5$
into the arbitrary point $\left( V^1,V^2,...,V^5\right) \in
\Sigma _\rho ^5\subset \mathcal{R}^5$ with curvature $\rho,$
$(V_A V^A=-\rho ^2, V^A=t^A\rho ).$ Matrix $b$ can be expressed as
\begin{equation*}
b=\left(
\begin{array}{cc}
\delta _{\quad \underline{\beta }}^{\underline{\alpha }}+\frac{t^{\underline{%
\alpha }}t_{\underline{\beta }}}{\left( 1+t^5\right) } & t^{\underline{%
\alpha }} \\
t_{\underline{\beta }} & t^5%
\end{array}
\right) .
\end{equation*}

The de Sitter gauge field is associated with a
${\mathit{s}o}_{\left( \eta \right) }\left( 5\right) $--valued
connection 1--form
\begin{equation}  \label{dspot}
\widetilde{\Omega }=\left(
\begin{array}{cc}
\omega _{\quad \underline{\beta }}^{\underline{\alpha }} &
\widetilde{\theta
}^{\underline{\alpha }} \\
\widetilde{\theta }_{\underline{\beta }} & 0%
\end{array}
\right) ,
\end{equation}
where $\omega _{\quad \underline{\beta }}^{\underline{\alpha }}\in
so(4)_{\left( \eta \right) },$ $\widetilde{\theta
}^{\underline{\alpha }}\in
\mathcal{R}^4,\widetilde{\theta }_{\underline{\beta }}\in \eta _{\underline{%
\beta }\underline{\alpha }}\widetilde{\theta }^{\underline{\alpha
}}.$

Because $S_{\left( \eta \right) }$--transforms mix the components
of the
matrix $\omega _{\quad \underline{\beta }}^{\underline{\alpha }}$ and $%
\widetilde{\theta }^{\underline{\alpha }}$ fields in
(\ref{dspot}) (the introduced para\-met\-ri\-za\-ti\-on is
invariant on action on $SO_{\left( \eta \right) }\left( 4\right)
$ group we cannot identify $\omega _{\quad
\underline{\beta }}^{\underline{\alpha }}$ and $\widetilde{\theta }^{%
\underline{\alpha }},$ respectively, with the connection $\Gamma
_{~\beta \gamma }^{\alpha }$ and the fundamental form $\chi
^{\alpha }$ in a metric--affine spacetime). To avoid this
difficulty we consider \cite{ts} a nonlinear gauge realization of
the de Sitter group $S_{\left( \eta \right) }, $ namely, we
introduce into consideration the nonlinear gauge field
\begin{equation}
\Gamma =b^{-1}{\widetilde{\Omega }}b+b^{-1}db=\left(
\begin{array}{cc}
\Gamma _{~\underline{\beta }}^{\underline{\alpha }} & \theta ^{\underline{%
\alpha }} \\
\theta _{\underline{\beta }} & 0%
\end{array}%
\right) ,  \label{npot}
\end{equation}%
where
\begin{eqnarray}
\Gamma _{\quad \underline{\beta }}^{\underline{\alpha }}
&=&\omega _{\quad
\underline{\beta }}^{\underline{\alpha }}-\left( t^{\underline{\alpha }}Dt_{%
\underline{\beta }}-t_{\underline{\beta }}Dt^{\underline{\alpha
}}\right)
/\left( 1+t^{5}\right) ,  \notag \\
\theta ^{\underline{\alpha }} &=&t^{5}\widetilde{\theta
}^{\underline{\alpha
}}+Dt^{\underline{\alpha }}-t^{\underline{\alpha }}\left( dt^{5}+\widetilde{%
\theta }_{\underline{\gamma }}t^{\underline{\gamma }}\right)
/\left(
1+t^{5}\right) ,  \notag \\
Dt^{\underline{\alpha }} &=&dt^{\underline{\alpha }}+\omega
_{\quad \underline{\beta }}^{\underline{\alpha
}}t^{\underline{\beta }}.  \notag
\end{eqnarray}

The action of the group $S\left( \eta \right) $ is nonlinear,
yielding transforms
\begin{equation*}
\Gamma ^{\prime }=L^{\prime }\Gamma \left( L^{\prime }\right)
^{-1}+L^{\prime }d\left( L^{\prime }\right) ^{-1},\theta ^{\prime
}=L\theta ,
\end{equation*}%
where the nonlinear matrix--valued function
\begin{equation*}
L^{\prime }=L^{\prime }\left( t^{\alpha },b,B_{T}\right)
\end{equation*}%
is defined from $B_{b}=b^{\prime }B_{L^{\prime }}$ (see the parametrization (%
\ref{parametriz})). The de Sitter algebra with generators
(\ref{dsca}) and
nonlinear gauge transforms of type (\ref{npot}) is denoted $\mathcal{A}%
_{I}^{(dS)}.$

\subsubsection{\ De Sitter nonlinear gauge gravity and Einstein and Finsler
like gravity}

Let us consider the de Sitter nonlinear gauge gravitational connection (\ref%
{npot}) rewritten in the form
\begin{equation}
\Gamma =\left(
\begin{array}{cc}
\Gamma _{\quad \underline{\beta }}^{\underline{\alpha }} & l_{0}^{-1}\chi ^{%
\underline{\alpha }} \\
l_{0}^{-1}\chi _{\underline{\beta }} & 0%
\end{array}%
\right)  \label{1a}
\end{equation}%
where
\begin{eqnarray}
\Gamma _{\quad \underline{\beta }}^{\underline{\alpha }}
&=&\Gamma _{\quad
\underline{\beta }\mu }^{\underline{\alpha }}\delta u^{\mu },  \notag \\
\Gamma _{\quad \underline{\beta }\mu }^{\underline{\alpha }}
&=&\chi _{\quad
\alpha }^{\underline{\alpha }}\chi _{\quad \beta }^{\underline{\beta }%
}\Gamma _{\quad \beta \gamma }^{\alpha }+\chi _{\quad \alpha }^{\underline{%
\alpha }}\delta _{\mu }\chi _{\quad \underline{\beta }}^{\alpha },\chi ^{%
\underline{\alpha }}=\chi _{\quad \mu }^{\underline{\alpha
}}\delta u^{\mu }, \notag
\end{eqnarray}%
and
\begin{equation*}
G_{\alpha \beta }=\chi _{\quad \alpha }^{\underline{\alpha }}\chi
_{\quad \beta }^{\underline{\beta }}\eta _{\underline{\alpha
}\underline{\beta }},
\end{equation*}%
$\eta _{\underline{\alpha }\underline{\beta }}=\left(
1,-1,...,-1\right) $ and $l_{0}$ is a dimensional constant. As
$\Gamma _{\quad \beta \gamma }^{\alpha }$ we take the Christoffel
symbols for the Einstein theory, or every type of d--connection
(\ref{dcon}) for an anisotropic spacetime. Correspondingly,
$G_{\alpha \beta }$ can be the pseudo--Rieamannian metric in
general relativity or any d--metric (\ref{dmetric}), which can be
particularized for the anholonomic Einstein gravity
(\ref{dmetrr}) \ or for a Finsler type gravity (\ref{dmetricf}).

The curvature of (\ref{1a}),
\begin{equation*}
\mathcal{R}^{(\Gamma )}=d\Gamma +\Gamma \bigwedge \Gamma ,
\end{equation*}%
can be written
\begin{equation}
\mathcal{R}^{(\Gamma )}=\left(
\begin{array}{cc}
\mathcal{R}_{\quad \underline{\beta }}^{\underline{\alpha }}+l_{0}^{-1}\pi _{%
\underline{\beta }}^{\underline{\alpha }} &
l_{0}^{-1}T^{\underline{\alpha }}
\\
l_{0}^{-1}T^{\underline{\beta }} & 0%
\end{array}%
\right) ,  \label{2a}
\end{equation}%
where
\begin{equation*}
\pi _{\underline{\beta }}^{\underline{\alpha }}=\chi ^{\underline{\alpha }%
}\bigwedge \chi _{\underline{\beta }},\mathcal{R}_{\quad \underline{\beta }%
}^{\underline{\alpha }}=\frac{1}{2}\mathcal{R}_{\quad
\underline{\beta }\mu \nu }^{\underline{\alpha }}\delta u^{\mu
}\bigwedge \delta u^{\nu },
\end{equation*}%
and
\begin{equation*}
\mathcal{R}_{\quad \underline{\beta }\mu \nu }^{\underline{\alpha }}=\chi _{%
\underline{\beta }}^{\quad \beta }\chi _{\alpha }^{\quad \underline{\alpha }%
}R_{\quad \beta _{\mu \nu }}^{\alpha }.
\end{equation*}%
with the $R_{\quad \beta {\mu \nu }}^{\alpha }$ being the
metric--affine (for Ein\-stein-\-Car\-tan-Weyl spaces), or the
(pseudo) Riemannian curvature, or for anisotropic spaces the
d--curvature (\ref{dcurvatures}). The de Sitter gauge group is
semisimple and we are able to construct a variational gauge
gravitational theory with the Lagrangian
\begin{equation*}
L=L_{\left( G\right) }+L_{\left( m\right) }
\end{equation*}%
where the gauge gravitational Lagrangian is defined
\begin{equation*}
L_{\left( G\right) }=\frac{1}{4\pi }Tr\left( \mathcal{R}^{(\Gamma
)}\bigwedge \ast _{G}\mathcal{R}^{(\Gamma )}\right)
=\mathcal{L}_{\left( G\right) }\left| G\right| ^{1/2}\delta
^{n+m}u,
\end{equation*}%
with
\begin{equation}
\mathcal{L}_{\left( G\right) }=\frac{1}{2l^{2}}T_{\quad \mu \nu }^{%
\underline{\alpha }}T_{\underline{\alpha }}^{\quad \mu \nu }+\frac{1}{%
8\lambda }\mathcal{R}_{\quad \underline{\beta }\mu \nu }^{\underline{\alpha }%
}\mathcal{R}_{\quad \underline{\alpha }}^{\underline{\beta }\quad
\mu \nu }{}-\frac{1}{l^{2}}\left( {\overleftarrow{R}}\left(
\Gamma \right) -2\lambda _{1}\right) ,  \notag
\end{equation}%
$\delta ^{4}u$ being the volume element, $T_{\quad \mu \nu }^{\underline{%
\alpha }}=\chi _{\quad \alpha }^{\underline{\alpha }}T_{\quad \mu
\nu
}^{\alpha }$ (the gravitational constant $l^{2}$ satisfies the relations $%
l^{2}=2l_{0}^{2}\lambda ,\lambda _{1}=-3/l_{0}),\quad Tr$ denotes
the trace on $\underline{\alpha },\underline{\beta }$ indices,
and the matter field Lagrangian is defined
\begin{equation*}
L_{\left( m\right) }=-1\frac{1}{2}Tr\left( \Gamma \bigwedge \ast _{G}%
\mathcal{I}\right) =\mathcal{L}_{\left( m\right) }\left| G\right|
^{1/2}\delta ^{n+m}u,
\end{equation*}%
where
\begin{equation*}
\mathcal{L}_{\left( m\right) }=\frac{1}{2}\Gamma _{\quad \underline{\beta }%
\mu }^{\underline{\alpha }}S_{\quad \alpha }^{\underline{\beta
}\quad \mu }-t_{\quad \underline{\alpha }}^{\mu }l_{\quad \mu
}^{\underline{\alpha }}.
\end{equation*}%
The matter field source $\mathcal{J}$ is obtained as a variational
derivation of $\mathcal{L}_{\left( m\right) }$ on $\Gamma $ and is
parametrized as
\begin{equation*}
\mathcal{J}=\left(
\begin{array}{cc}
S_{\quad \underline{\beta }}^{\underline{\alpha }} & -l_{0}t^{\underline{%
\alpha }} \\
-l_{0}t_{\underline{\beta }} & 0%
\end{array}%
\right)
\end{equation*}%
with $t^{\underline{\alpha }}=t_{\quad \mu }^{\underline{\alpha
}}\delta u^{\mu }$ and $S_{\quad \underline{\beta
}}^{\underline{\alpha }}=S_{\quad \underline{\beta }\mu
}^{\underline{\alpha }}\delta u^{\mu }$ being respectively the
canonical tensors of energy--momentum and spin density.

Varying the action
\begin{equation*}
S=\int \delta ^{4}u\left( \mathcal{L}_{\left( G\right)
}+\mathcal{L}_{\left( m\right) }\right)
\end{equation*}%
on the $\Gamma $--variables (1a), we obtain the
gau\-ge--gra\-vi\-ta\-ti\-on\-al field equations, in general,
with local anisotropy,
\begin{equation}
d\left( \ast \mathcal{R}^{(\Gamma )}\right) +\Gamma \bigwedge
\left( \ast \mathcal{R}^{(\Gamma )}\right) -\left( \ast
\mathcal{R}^{(\Gamma )}\right) \bigwedge \Gamma =-\lambda \left(
\ast \mathcal{J}\right) ,  \label{3a}
\end{equation}%
were the Hodge operator $\ast $ is used.

Specifying the variations on $\Gamma _{\quad \underline{\beta }}^{\underline{%
\alpha }}$ and $\chi $--variables, we rewrite (\ref{3a})
\begin{eqnarray}
\widehat{\mathcal{D}}\left( \ast \mathcal{R}^{(\Gamma )}\right) &&+\frac{%
2\lambda }{l^{2}}(\widehat{\mathcal{D}}\left( \ast \pi \right)
+\chi \bigwedge \left( \ast T^{T}\right) -\left( \ast T\right)
\bigwedge \chi
^{T})=-\lambda \left( \ast S\right) ,  \notag \\
\widehat{\mathcal{D}}\left( \ast T\right) &-&\left( \ast \mathcal{R}%
^{(\Gamma )}\right) \bigwedge \chi -\frac{2\lambda }{l^{2}}\left(
\ast \pi \right) \bigwedge \chi =\frac{l^{2}}{2}\left( \ast
t+\frac{1}{\lambda }\ast \tau \right) ,  \notag
\end{eqnarray}%
where
\begin{eqnarray}
T^{t} &=&\{T_{\underline{\alpha }}=\eta _{\underline{\alpha }\underline{%
\beta }}T^{\underline{\beta }},~T^{\underline{\beta
}}=\frac{1}{2}T_{\quad \mu \nu }^{\underline{\beta }}\delta
u^{\mu }\bigwedge \delta u^{\nu }\},
\notag \\
\chi ^{T} &=&\{\chi _{\underline{\alpha }}=\eta _{\underline{\alpha }%
\underline{\beta }}\chi ^{\underline{\beta }},~\chi ^{\underline{\beta }%
}=\chi _{\quad \mu }^{\widehat{\beta }}\delta u^{\mu }\},\qquad \widehat{%
\mathcal{D}}=d+\widehat{\Gamma },  \notag
\end{eqnarray}%
($\widehat{\Gamma }$ acts as $\Gamma _{\quad \underline{\beta }\mu }^{%
\underline{\alpha }}$ on indices $\underline{\gamma },\underline{\delta }%
,... $ and as $\Gamma _{\quad \beta \mu }^{\alpha }$ on indices
$\gamma ,\delta ,...).$ The value $\tau $ defines the
energy--momentum tensor of the
gauge gravitational field $\widehat{\Gamma }:$%
\begin{equation*}
\tau _{\mu \nu }\left( \widehat{\Gamma }\right)
=\frac{1}{2}Tr\left(
\mathcal{R}_{\mu \alpha }\mathcal{R}_{\quad \nu }^{\alpha }-\frac{1}{4}%
\mathcal{R}_{\alpha \beta }\mathcal{R}^{\alpha \beta }G_{\mu \nu
}\right) .
\end{equation*}

Equations (\ref{3a}) make up the complete system of variational
field equations for nonlinear de Sitter gauge anisotropic gravity.

We note that we can obtain a nonvariational Poincare gauge
gravitational theory if we consider the contraction of the gauge
potential (\ref{1a}) to a potential with values in the Poincare
Lie algebra
\begin{equation}
\Gamma =\left(
\begin{array}{cc}
\Gamma _{\quad \widehat{\beta }}^{\widehat{\alpha }} & l_{0}^{-1}\chi ^{%
\widehat{\alpha }} \\
l_{0}^{-1}\chi _{\widehat{\beta }} & 0%
\end{array}%
\right) \rightarrow \Gamma =\left(
\begin{array}{cc}
\Gamma _{\quad \widehat{\beta }}^{\widehat{\alpha }} & l_{0}^{-1}\chi ^{%
\widehat{\alpha }} \\
0 & 0%
\end{array}%
\right) .  \label{4a}
\end{equation}%
A similar gauge potential was considered in the formalism of
linear and affine frame bundles on curved spacetimes by Popov and
Daikhin \cite{pd}. They treated (\ref{4a}) as the Cartan
connection form for affine gauge like gravity and by using 'pure'
geometric methods proved that the Yang--Mills equations of their
theory are equivalent, after projection on the base, to the
Einstein equations. The main conclusion for a such approach to
Einstein gravity is that this theory admits an equivalent
formulation as a gauge model but with a nonsemisimple structural
gauge group. In order to have a variational theory on the total
bundle space it is necessary to introduce an auxiliary bilinear
form on the typical fiber, instead of degenerated Killing form;
the coefficients of auxiliary form disappear after pojection on
the base. An alternative variant is to consider a gauge
gravitational theory when the gauge group was minimally extended
to the de\ Sitter one with nondegenerated Killing form. The
nonlinear realizations have to be introduced if we consider in a
common fashion both the frame (tetradic) and
connection components included as the coefficients of the potential (\ref{1a}%
). Finally, we note that the models of de Sitter gauge gravity
were generalized for Finsler and Lagrange theories in Refs.
\cite{vgauge,vmon1}.

\subsubsection{Enveloping nonlinear de Sitter algebra valued connection}

Let now us consider a noncommutative space. In this case the
gauge fields are elements of the algebra $\widehat{\psi }\in
\mathcal{A}_{I}^{(dS)}$ that
form the nonlinear representation of the de Sitter algebra ${\mathit{s}o}%
_{\left( \eta \right) }\left( 5\right) $ when the whole algebra is denoted $%
\mathcal{A}_{z}^{(dS)}.$ Under a nonlinear de Sitter
transformation the
elements transform as follows%
\begin{equation*}
\delta \widehat{\psi }=i\widehat{\gamma }\widehat{\psi
},\widehat{\psi }\in \mathcal{A}_{u},\widehat{\gamma }\in
\mathcal{A}_{z}^{(dS)}.
\end{equation*}%
So, the action of the generators (\ref{dsca}) on $\widehat{\psi
}$ is
defined as this element is supposed to form a nonlinear representation of $%
\mathcal{A}_{I}^{(dS)}$ and, in consequence, $\delta
\widehat{\psi }\in \mathcal{A}_{u}$ despite $\widehat{\gamma }\in
\mathcal{A}_{z}^{(dS)}.$ It
should be emphasized that independent of a representation the object $%
\widehat{\gamma }$ takes values in enveloping de Sitter algebra
and not in a Lie algebra as would be for commuting spaces. The
same holds for the connections that we introduce \cite{mssw}, in
order to define covariant
coordinates,%
\begin{equation*}
\widehat{U}^{\nu }=\widehat{u}^{v}+\widehat{\Gamma }^{\nu },\widehat{\Gamma }%
^{\nu }\in \mathcal{A}_{z}^{(dS)}.
\end{equation*}

The values $\widehat{U}^{\nu }\widehat{\psi }$ transform
covariantly,
\begin{equation*}
\delta \widehat{U}^{\nu }\widehat{\psi }=i\widehat{\gamma }\widehat{U}^{\nu }%
\widehat{\psi },
\end{equation*}%
if and only if the connection $\widehat{\Gamma }^{\nu }$
satisfies the
transformation law of the enveloping nonlinear realized de Sitter algebra,%
\begin{equation*}
\delta \widehat{\Gamma }^{\nu }\widehat{\psi }=-i[\widehat{u}^{v},\widehat{%
\gamma }]+i[\widehat{\gamma },\widehat{\Gamma }^{\nu }],
\end{equation*}%
where $\delta \widehat{\Gamma }^{\nu }\in
\mathcal{A}_{z}^{(dS)}.$ The enveloping algebra--valued
connection has infinitely many component fields. Nevertheless, it
was shown that all the component fields can be induced from
a Lie algebra--valued connection by a Seiberg--Witten map (\cite{sw,js} and %
\cite{bsst} for $SO(n)$ and $Sp(n)).$ In this subsection we show
that similar constructions could be proposed for nonlinear
realizations of de
Sitter algebra when the transformation of the connection is considered%
\begin{equation*}
\delta \widehat{\Gamma }^{\nu }=-i[u^{\nu },^{\ast }~\widehat{\gamma }]+i[%
\widehat{\gamma },^{\ast }~\widehat{\Gamma }^{\nu }].
\end{equation*}%
For simplicity, we treat in more detail the canonical case with
the star
product (\ref{csp1}). The first term in the variation $\delta \widehat{%
\Gamma }^{\nu }$ gives
\begin{equation*}
-i[u^{\nu },^{\ast }~\widehat{\gamma }]=\theta ^{\nu \mu }\frac{\partial }{%
\partial u^{\mu }}\gamma .
\end{equation*}%
Assuming that the variation of $\widehat{\Gamma }^{\nu }=\theta
^{\nu \mu
}Q_{\mu }$ starts with a linear term in $\theta ,$ we have%
\begin{equation*}
\delta \widehat{\Gamma }^{\nu }=\theta ^{\nu \mu }\delta Q_{\mu
},~\delta
Q_{\mu }=\frac{\partial }{\partial u^{\mu }}\gamma +i[\widehat{\gamma }%
,^{\ast }~Q_{\mu }].
\end{equation*}%
We follow the method of calculation from the papers
\cite{mssw,jssw} and expand the star product (\ref{csp1}) in
$\theta $ but not in $g_{a}$ and
find to first order in $\theta ,$%
\begin{equation}
\gamma =\gamma _{\underline{a}}^{1}I^{\underline{a}}+\gamma _{\underline{a}%
\underline{b}}^{1}I^{\underline{a}}I^{\underline{b}}+...,Q_{\mu }=q_{\mu ,%
\underline{a}}^{1}I^{\underline{a}}+q_{\mu ,\underline{a}\underline{b}%
}^{2}I^{\underline{a}}I^{\underline{b}}+...  \label{seriesa}
\end{equation}%
where $\gamma _{\underline{a}}^{1}$ and $q_{\mu
,\underline{a}}^{1}$ are of
order zero in $\theta $ and $\gamma _{\underline{a}\underline{b}}^{1}$ and $%
q_{\mu ,\underline{a}\underline{b}}^{2}$ are of second order in
$\theta .$ The expansion in $I^{\underline{b}}$ leads to an
expansion in $g_{a}$ of the $\ast $--product because the higher
order $I^{\underline{b}}$--derivatives
vanish. For de Sitter case as $I^{\underline{b}}$ we take the generators (%
\ref{dsca}), see commutators (\ref{commutators1}), with the
corresponding de
Sitter structure constants $f_{~\underline{d}}^{\underline{b}\underline{c}%
}\simeq f_{~\underline{\beta }}^{\underline{\alpha
}\underline{\beta }}$ (in our further identifications with
spacetime objects like frames and connections we shall use Greek
indices).

The result of calculation of variations of (\ref{seriesa}), by
using $g_{a}$
to the order given in (\ref{gdecomp}), is%
\begin{eqnarray}
\delta q_{\mu ,\underline{a}}^{1} &=&\frac{\partial \gamma _{\underline{a}%
}^{1}}{\partial u^{\mu }}-f_{~\underline{a}}^{\underline{b}\underline{c}%
}\gamma _{\underline{b}}^{1}q_{\mu ,\underline{c}}^{1},  \notag \\
\delta Q_{\tau } &=&\theta ^{\mu \nu }\partial _{\mu }\gamma _{\underline{a}%
}^{1}\partial _{\nu }q_{\tau ,\underline{b}}^{1}I^{\underline{a}}I^{%
\underline{b}}+...,  \notag \\
\delta q_{\mu ,\underline{a}\underline{b}}^{2} &=&\partial _{\mu }\gamma _{%
\underline{a}\underline{b}}^{2}-\theta ^{\nu \tau }\partial _{\nu }\gamma _{%
\underline{a}}^{1}\partial _{\tau }q_{\mu ,\underline{b}}^{1}-2f_{~%
\underline{a}}^{\underline{b}\underline{c}}\{\gamma _{\underline{b}%
}^{1}q_{\mu ,\underline{c}\underline{d}}^{2}+\gamma _{\underline{b}%
\underline{d}}^{2}q_{\mu ,\underline{c}}^{1}\}.  \notag
\end{eqnarray}

Next, we introduce the objects $\varepsilon ,$ taking the values
in de
Sitter Lie algebra and $W_{\mu },$ being enveloping de Sitter algebra valued,%
\begin{equation*}
\varepsilon =\gamma _{\underline{a}}^{1}I^{\underline{a}}\mbox{
and
}W_{\mu }=q_{\mu ,\underline{a}\underline{b}}^{2}I^{\underline{a}}I^{%
\underline{b}},
\end{equation*}%
with the variation $\delta W_{\mu }$ satisfying the equation
\cite{mssw,jssw}
\begin{equation*}
\delta W_{\mu }=\partial _{\mu }(\gamma _{\underline{a}\underline{b}}^{2}I^{%
\underline{a}}I^{\underline{b}})-\frac{1}{2}\theta ^{\tau \lambda
}\{\partial _{\tau }\varepsilon ,\partial _{\lambda }q_{\mu
}\}+{}i[\varepsilon ,W_{\mu }]+i[(\gamma _{\underline{a}\underline{b}}^{2}I^{%
\underline{a}}I^{\underline{b}}),q_{\nu }].
\end{equation*}%
This equation has the solution (found in \cite{mssw,sw})%
\begin{equation}
\gamma _{\underline{a}\underline{b}}^{2}=\frac{1}{2}\theta ^{\nu
\mu
}(\partial _{\nu }\gamma _{\underline{a}}^{1})q_{\mu ,\underline{b}%
}^{1},~q_{\mu ,\underline{a}\underline{b}}^{2}=-\frac{1}{2}\theta
^{\nu \tau
}q_{\nu ,\underline{a}}^{1}\left( \partial _{\tau }q_{\mu ,\underline{b}%
}^{1}+R_{\tau \mu ,\underline{b}}^{1}\right)  \notag
\end{equation}%
where
\begin{equation*}
R_{\tau \mu ,\underline{b}}^{1}=\partial _{\tau }q_{\mu ,\underline{b}%
}^{1}-\partial _{\mu }q_{\tau ,\underline{b}}^{1}+f_{~\underline{d}}^{%
\underline{e}\underline{c}}q_{\tau ,\underline{e}}^{1}q_{\mu ,\underline{e}%
}^{1}
\end{equation*}%
could be identified with the coefficients $\mathcal{R}_{\quad \underline{%
\beta }\mu \nu }^{\underline{\alpha }}$ of de Sitter nonlinear
gauge gravity
curvature (see formula (\ref{2a})) if in the commutative limit $q_{\mu ,%
\underline{b}}^{1}\simeq \left(
\begin{array}{cc}
\Gamma _{\quad \underline{\beta }}^{\underline{\alpha }} & l_{0}^{-1}\chi ^{%
\underline{\alpha }} \\
l_{0}^{-1}\chi _{\underline{\beta }} & 0%
\end{array}%
\right) $ (see (\ref{1a})).

The below presented procedure can be generalized to all the
higher powers of $\theta $.

\subsection{Noncommutative Gravity Covariant Gauge Dynamics}

\subsubsection{First order corrections to gravitational curvature}

The constructions from the previous section are summarized by the
conclusion
that the de Sitter algebra valued object $\varepsilon =\gamma _{\underline{a}%
}^1\left( u\right) I^{\underline{a}}$ determines all the terms in
the
enveloping algebra%
\begin{equation*}
\gamma =\gamma _{\underline{a}}^1I^{\underline{a}}+\frac 14\theta
^{\nu \mu
}\partial _\nu \gamma _{\underline{a}}^1\ q_{\mu ,\underline{b}}^1\left( I^{%
\underline{a}}I^{\underline{b}}+I^{\underline{b}}I^{\underline{a}}\right)
+...
\end{equation*}
and the gauge transformations are defined by $\gamma _{\underline{a}%
}^1\left( u\right) $ and $q_{\mu ,\underline{b}}^1(u),$ when
\begin{equation*}
\delta _{\gamma ^1}\psi =i\gamma \left( \gamma ^1,q_\mu ^1\right)
*\psi .
\end{equation*}
For de Sitter enveloping algebras one holds the general formula
for
compositions of two transformations%
\begin{equation*}
\delta _\gamma \delta _\varsigma -\delta _\varsigma \delta
_\gamma =\delta _{i(\varsigma *\gamma -\gamma *\varsigma )}
\end{equation*}
which is also true for the restricted transformations defined by $\gamma ^1,$%
\begin{equation*}
\delta _{\gamma ^1}\delta _{\varsigma ^1}-\delta _{\varsigma
^1}\delta _{\gamma ^1}=\delta _{i(\varsigma ^1*\gamma ^1-\gamma
^1*\varsigma ^1)}.
\end{equation*}

Applying the formula (\ref{csp1}) we calculate%
\begin{eqnarray*}
\lbrack \gamma ,^{\ast }\zeta ] &=&i\gamma _{\underline{a}}^{1}\zeta _{%
\underline{b}}^{1}f_{~\underline{c}}^{\underline{a}\underline{b}}I^{%
\underline{c}}+\frac{i}{2}\theta ^{\nu \mu }\{\partial _{v}\left( \gamma _{%
\underline{a}}^{1}\zeta _{\underline{b}}^{1}f_{~\underline{c}}^{\underline{a}%
\underline{b}}\right) q_{\mu ,\underline{c}} \\
&&+{}\left( \gamma _{\underline{a}}^{1}\partial _{v}\zeta _{\underline{b}%
}^{1}-\zeta _{\underline{a}}^{1}\partial _{v}\gamma _{\underline{b}%
}^{1}\right) q_{\mu ,\underline{b}}f_{~\underline{c}}^{\underline{a}%
\underline{b}}+2\partial _{v}\gamma _{\underline{a}}^{1}\partial
_{\mu }\zeta
_{\underline{b}}^{1}\}I^{\underline{d}}I^{\underline{c}}.
\end{eqnarray*}%
Such commutators could be used for definition of tensors
\cite{mssw}
\begin{equation}
\widehat{S}^{\mu \nu }=[\widehat{U}^{\mu },\widehat{U}^{\nu }]-i\widehat{%
\theta }^{\mu \nu },  \label{tensor1}
\end{equation}%
where $\widehat{\theta }^{\mu \nu }$ is respectively stated for
the canonical, Lie and quantum plane structures. Under the
general enveloping
algebra one holds the transform%
\begin{equation*}
\delta \widehat{S}^{\mu \nu }=i[\widehat{\gamma
},\widehat{S}^{\mu \nu }].
\end{equation*}%
For instance, the canonical case is characterized by%
\begin{eqnarray}
S^{\mu \nu } &=&i\theta ^{\mu \tau }\partial _{\tau }\Gamma ^{\nu
}-i\theta ^{\nu \tau }\partial _{\tau }\Gamma ^{\mu }+\Gamma
^{\mu }\ast \Gamma ^{\nu
}-\Gamma ^{\nu }\ast \Gamma ^{\mu }  \notag \\
&=&\theta ^{\mu \tau }\theta ^{\nu \lambda }\{\partial _{\tau
}Q_{\lambda }-\partial _{\lambda }Q_{\tau }+Q_{\tau }\ast
Q_{\lambda }-Q_{\lambda }\ast Q_{\tau }\}.  \notag
\end{eqnarray}%
By introducing the gravitational gauge strength (curvature)
\begin{equation}
R_{\tau \lambda }=\partial _{\tau }Q_{\lambda }-\partial
_{\lambda }Q_{\tau }+Q_{\tau }\ast Q_{\lambda }-Q_{\lambda }\ast
Q_{\tau },  \label{qcurv}
\end{equation}%
which could be treated as a noncommutative extension of de Sitter
nonlinear gauge gravitational curvature (2a), we calculate
\begin{equation}
R_{\tau \lambda ,\underline{a}}=R_{\tau \lambda
,\underline{a}}^{1}+\theta
^{\mu \nu }\{R_{\tau \mu ,\underline{a}}^{1}R_{\lambda \nu ,\underline{b}%
}^{1}{}-\frac{1}{2}q_{\mu ,\underline{a}}^{1}\left[ (D_{\nu
}R_{\tau \lambda
,\underline{b}}^{1})+\partial _{\nu }R_{\tau \lambda ,\underline{b}}^{1}%
\right] \}I^{\underline{b}},  \notag
\end{equation}%
where the gauge gravitation covariant derivative is introduced,%
\begin{equation*}
(D_{\nu }R_{\tau \lambda ,\underline{b}}^{1})=\partial _{\nu
}R_{\tau
\lambda ,\underline{b}}^{1}+q_{\nu ,\underline{c}}R_{\tau \lambda ,%
\underline{d}}^{1}f_{~\underline{b}}^{\underline{c}\underline{d}}.
\end{equation*}%
Following the gauge transformation laws for $\gamma $ and $q^{1}$
we find
\begin{equation*}
\delta _{\gamma ^{1}}R_{\tau \lambda }^{1}=i\left[ \gamma ,^{\ast
}R_{\tau \lambda }^{1}\right]
\end{equation*}%
with the restricted form of $\gamma .$

Such formulas were proved in references \cite{sw} for usual gauge
(nongravitational) fields. Here we reconsidered them for
gravitational gauge fields.

\subsubsection{Gauge covariant gravitational dynamics}

Following the nonlinear realization of de Sitter algebra and the $\ast $%
--formalism we can formulate a dynamics of noncommutative spaces.
Derivatives can be introduced in such a way that one does not
obtain new relations for the coordinates. In this case a Leibniz
rule can be defined that
\begin{equation*}
\widehat{\partial }_{\mu }\widehat{u}^{\nu }=\delta _{\mu }^{\nu
}+d_{\mu \sigma }^{\nu \tau }\ \widehat{u}^{\sigma }\
\widehat{\partial }_{\tau }
\end{equation*}%
where the coefficients $d_{\mu \sigma }^{\nu \tau }=\delta
_{\sigma }^{\nu
}\delta _{\mu }^{\tau }$ are chosen to have not new relations when $\widehat{%
\partial }_{\mu }$ acts again to the right hand side. In consequence one
holds the $\ast $--derivative formulas
\begin{equation}
{\partial }_{\tau }\ast f=\frac{\partial }{\partial u^{\tau }}f+f\ast {%
\partial }_{\tau },~[{\partial }_{l},{{}^{\ast }}(f\ast g)]=([{\partial }%
_{l},{{}^{\ast }}f])\ast g+f\ast ([{\partial }_{l},{}^{\ast }g])
\notag
\end{equation}%
and the Stokes theorem%
\begin{equation*}
\int [\partial _{l},f]=\int d^{N}u[\partial _{l},^{\ast }f]=\int d^{N}u\frac{%
\partial }{\partial u^{l}}f=0,
\end{equation*}%
where, for the canonical structure, the integral is defined,%
\begin{equation*}
\int \widehat{f}=\int d^{N}uf\left( u^{1},...,u^{N}\right) .
\end{equation*}

An action can be introduced by using such integrals. For
instance, for a
tensor of type (\ref{tensor1}), when%
\begin{equation*}
\delta \widehat{L}=i\left[ \widehat{\gamma },\widehat{L}\right] ,
\end{equation*}
we can define a gauge invariant action%
\begin{equation*}
W=\int d^Nu\ Tr\widehat{L},~\delta W=0,
\end{equation*}
were the trace has to be taken for the group generators.

For the nonlinear de Sitter gauge gravity a proper action is
\begin{equation*}
L=\frac{1}{4}R_{\tau \lambda }R^{\tau \lambda },
\end{equation*}%
where $R_{\tau \lambda }$ is defined by (\ref{qcurv}) (in the
commutative limit we shall obtain the connection (\ref{1a})). In
this case the dynamic of noncommutative space is entirely
formulated in the framework of quantum field theory of gauge
fields. In general, we are dealing with anisotropic gauge
gravitational interactions. The method works for matter fields as
well to restrictions to the general relativity theory.

\section{Outlook and Conclusions}

In this work we have extended the A. Connes' approach to
noncommutative geometry by introducing into consideration
anholonomic frames and locally anisotropic structures. We defined
nonlinear connections for finite projective module spaces
(noncommutative generalization of vector bundles) and related
this geometry with the E. Cartan's moving frame method.

We have explicitly shown that the functional analytic approach and
noncommutative $C^{\ast }$--algebras may be transformed into
arena of modeling geometries and physical theories with generic
local anisotropy, for instance, the anholonomic Riemannian
gravity and generalized Finsler like geometries. The formalism of
spectral triples elaborated for vector bundles provided with
nonlinear connection structure allows a functional and algebraic
generation of new types of anholonomic/ anisotropic interactions.

A novel future in our work is that by applying anholonomic
transforms associated to some nonlinear connections we may
generate various type of spinor, gauge and gravity models,
subjected to some anholonomic constraints and/or with generic
anisotropic interactions, which can be included in noncommutative
field theory.

We can address a number of questions which were put or solved
partially in this paper and may have further generalizations:

One of the question is how to combine the noncommutative geometry
contained in string theory with locally anisotropic
configurations arising in the low energy limits. It is known that
the nonsymmetric background field results in effective
noncommutative coordinates. In other turn, a (super) frame set
consisting from mixed subsets of holonomic and anholonomic
vectors may result in an anholonomic geometry with associated
nonlinear connection structure. A further work is to investigate
the conditions when from a string theory one appears explicit
variants of commutative--anisotropic, commutative--isotropic,
noncommutative--isotropic and, finally,
nocommutative--an\-iso\-tro\-pic geometries.

A second question is connected with the problem of definition of
noncommutative (pseudo)\ Riemannian metric structures which is
connected with nonsymmetric and/or complex metrics. We have
elaborated variants of noncommutative gauge gravity with
noncommutative representations of the affine and de Sitter
algebras which contains in the commutative limit an Yang--Mills
theory (with nonsemisimple structure group) being equivalent to
the Einstein theory. The gauge connection in such theories is
constructed from the frame and linear connection coefficients.
Metrics, in this case, arise as some effective configurations
which avoid problems with their noncommutative definition. The
approach can be generated as to include anholonomic frames and,
in consequence, to define anisotropic variants of commutative and
noncommutative gauge gravity with the Einstein type or Finsler
generalizations.

Another interesting open question is to establish a relation
between quantum groups and geometries with anisotropic models of
gravity and field theories. Different variants of quantum
generalizations for anholonomic frames with associated nonlinear
connection structures are possible.

Finally, we give some historical remarks. An approach to Finsler
and spinor like spaces of infinite dimensions (in Banach and/or
Hilbert spaces) and to nonsymmetric locally anisotropic metrics
was proposed by some authors belonging to the Romannian school on
Finsler geometry and generalizations (see, Refs.
\cite{ma,hrimiuc,anastasiei,atanasiu}). It could not be finalized
before ellaboration of the\ A. Connes' models of noncommutative
geometry and gravity and before  definition of Clifford and spinor
distinguished structures \cite{vspinors,vmon2},  formulation of
supersymmetric variants of Finsler spaces \cite{vsuper} and
establishing theirs relation to string theory
\cite{vstring,vstr2,vmon1}. This paper concludes a noncommutative
interference and a development of the mentioned results.

\subsection*{Acknowledgements}

~~ The work is supported by a NATO/Portugal fellowship at CENTRA,
Instituto Superior Tecnico, Lisbon. The author is grateful to R.
Ablamowicz, John Ryan and B. Fauser for collaboration and support
of his participation at ''The 6th International Conference on
Clifford Algebras'', Cookeville, Tennessee, USA (May, 20-25,
2002). He would like to thank S. Majid and C. Sochichiu for
usefull discussions and J. P. S. Lemos and M. Anastasiei for
hospitality and support.


\end{document}